\documentclass[%
 reprint,
 amsmath,amssymb,
 aps,
]{revtex4-2}
\usepackage[T2A]{fontenc}
\usepackage{graphicx}
\usepackage{dcolumn}
\usepackage{bm}
\usepackage{amssymb}
\usepackage{epstopdf}
\usepackage{booktabs}
\newcommand{\scs}{\scriptscriptstyle}
\newcommand{\smallm}{\scs (\!-\!)}
\newcommand{\smallp}{\scs (\!+\!)}

\newcommand{\smallpm}{\scs (\!\pm\!)}

\begin{document}

\preprint{APS/123-QED}

\title{Photonic bound states in the continuum governed by heating}

\author{A.I.~Krasnov}
\affiliation{Kirensky Institute of Physics, Federal Research Center KSC SB
RAS, Krasnoyarsk, 660036 Russia}
\affiliation{Siberian Federal University, Krasnoyarsk, 660041 Russia}%

\author{P. S. Pankin}
\email{pavel-s-pankin@iph.krasn.ru}
\affiliation{Kirensky Institute of Physics, Federal Research Center KSC SB RAS, Krasnoyarsk, 660036 Russia}
\affiliation{Siberian Federal University, Krasnoyarsk, 660041 Russia}%

\author{G. A. Romanenko}
\affiliation{Krasnoyarsk Scientific Center, Siberian Branch, Russian Academy of Sciences, Krasnoyarsk, 660036 Russia}
\affiliation{Kirensky Institute of Physics, Federal Research Center KSC SB
RAS, Krasnoyarsk, 660036 Russia}
\affiliation{Siberian State University of Science and Technology, Krasnoyarsk, 660037 Russia}

\author{V. S. Sutormin}
\affiliation{Kirensky Institute of Physics, Federal Research Center KSC SB
RAS, Krasnoyarsk, 660036 Russia}
\affiliation{Siberian Federal University, Krasnoyarsk, 660041 Russia}%

\author{D. N. Maksimov}
\affiliation{Kirensky Institute of Physics, Federal Research Center KSC SB
RAS, Krasnoyarsk, 660036 Russia}
\affiliation{Siberian Federal University, Krasnoyarsk, 660041 Russia}%

\author{S. Ya. Vetrov}
\affiliation{Siberian Federal University, Krasnoyarsk, 660041 Russia}%
\affiliation{Kirensky Institute of Physics, Federal Research Center KSC SB
RAS, Krasnoyarsk, 660036 Russia}

\author{I. V. Timofeev}
\affiliation{Kirensky Institute of Physics, Federal Research Center KSC SB
RAS, Krasnoyarsk, 660036 Russia}
\affiliation{Siberian Federal University, Krasnoyarsk, 660041 Russia}%

\altaffiliation[$\dagger$]{These authors equally contributed to this work}
\altaffiliation[*]{Corresponding author: pavel-s-pankin@iph.krasn.ru}


\date{\today}

\begin{abstract}
A photonic crystal microcavity with the liquid crystal resonant layer tunable by heating has been implemented. The multiple vanishing resonant lines corresponding to optical bound states in the continuum are observed. The abrupt behaviour of the resonant linewidth near the vanishing point can be used for temperature sensing. 

\end{abstract}

\maketitle


\section{\label{sec:level1}\protect introduction}

One-dimensional photonic crystal (PhC) is a periodic structure formed by layers with different refractive indices (RIs)~\cite{Joannopoulos2008bk}.
The optical thicknesses of the alternating layers are comparable with the wavelength, which leads to the Bragg diffraction of light.
In the photonic bandgap (PBG) spectral region the PhC reflects light with small losses.
The resonant layer embedded between two PhC mirrors forms a microcavity, which supports microcavity (MC) modes~\cite{kavokin2017microcavities}.

When the resonant layer is an anisotropic material
the MC mode can be transformed into a bound state in the continuum (BIC) \cite{Timofeev2018_BIC, Pankin2020BIC, Pankin2020Fano, Ignatyeva2020BIC, PankinLPR2021_Temp, Pankin_JOSAB2022, nabol2022fabryperot, KrasnovBIC2023}. 
The BIC is a nonradiative localized eigenmode ebmedded in the continuum of propagating waves~\cite{HsuChiaWei2016}. 
The BIC is a general wave phenomenon, which occurs in quantum mechanics, radio physics, photonics, and acoustics \cite{HsuChiaWei2016, sadreev2021interference, Koshelev2023, Azzam2020BICReview, Joseph2021}.
Variation of parameters of the system near the BIC affects the coupling between the localized mode and the continuum of propagating waves and thereby tunes the radiation decay rate. 
The BIC has been used in various applications, such as, lasers~\cite{Kodigala17, Kivshar_2020_Science_BIC_Laser, Pankin_LPR2021, huang2020vortex}, waveguides~\cite{Hsu2013, Bezus2018_BIC, GomisBresco_Torner2017_BIC1D, Mukherjee2021, Bulgakov2022metalwaveguide}, nanocavities~\cite{Rybin2017, Koshelev_nolinear2020}, amplified chiral response~\cite{shi2022chiral, gorkunov2020metasurfaces},  trapping and sorting of nanoparticles~\cite{Sadreev2022trapping}, perfect absorbers~\cite{wang2023plasmonicBIC, bikbaev2023absBIC}, etc.
Due to the narrow spectral line the quasi-BICs have been used for RI sensing~\cite{romano2020ultrasensitive, Maksimov2020sensor, Yusupov2021, Fedyanin2022sensor, Romano2022_BICSensor}, mechanical pressure sensing~\cite{Sadreev2022PRB} as well as for temperature sensing~\cite{Maksimov2022, wu2019giantGH_BIC, Yusupov2020tempBICsensor}. In this work, we demonstrate an optical BIC in a PhC microcavity tunable by heating the anisotropic liquid crystal (LC) resonant layer~\cite{Arkhipkin2008, Pankin2021APL, buzin2023hybrid} having in mind potential applications for temperature sensing.

\section{Model}

\begin{figure}[!ht]
\includegraphics{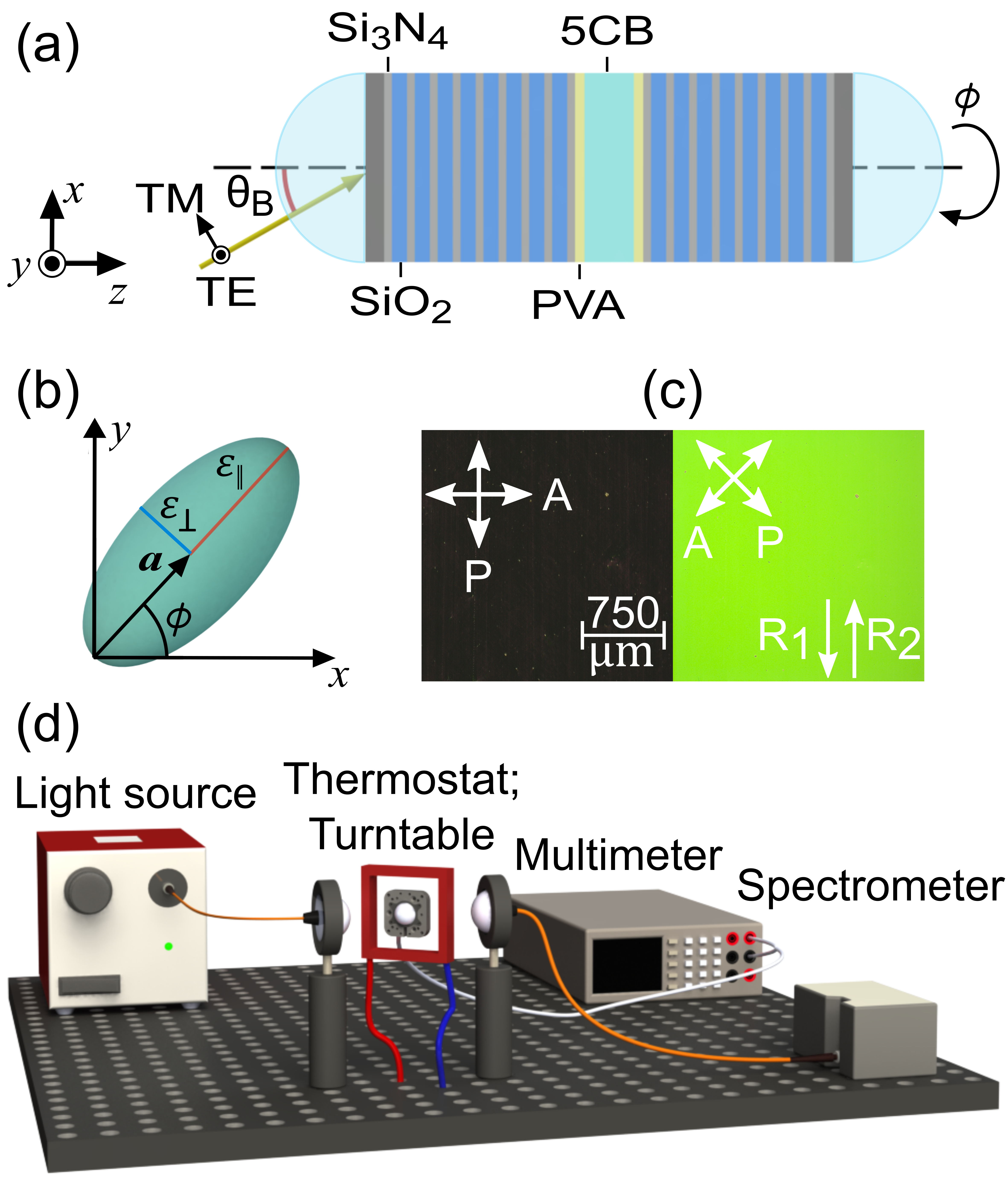}
\caption{(a) The microcavity model. (b) The orientation of the LC permittivity ellipsoid. 
(c) Polarizing optical microscope images of the LC layer texture taken in crossed polarizers. R$_\text{1}$ and R$_\text{2}$ are the PVA rubbing directions. Crossed double arrows show the direction of the polarizer (P) and analyzer (A). 
(d) The experimental set-up.}
\label{fig1}
\end{figure}

The microcavity consists of a LC layer embedded between two identical 1D PhCs, see Fig.~\ref{fig1}(a).
The silicon nitride (Si$_3$N$_4$) and silicon dioxide (SiO$_2$) layers are deposited on glass substrate by using the plasma enhanced chemical vapor deposition method. 
The number of periods in PhC is 8 plus one unpaired layer of silicon nitride. 
Polyvinyl alcohol (PVA) layers  are formed on each PhC by the spin-coating method and then mechanically rubbed to ensure a homogeneous planar alignment of the LC.
The gap between PhC mirrors is provided with Teflon spacers.
The 4-pentyl-4'-cyanobiphenyl (5CB) nematic LC is embedded into the gap by the capillary method.

The thicknesses and RIs of all layers are presented in Table \ref{Table_1}.
In nematic LCs the optical axis (OA) coincides with the 
preferred alignment of the long axes of the LC molecules which is described by the unit vector $\bm{a}~=~\left[\cos{(\phi)},~\sin{(\phi)},~0 \right]$, called the director, see Fig.~\ref{fig1}(b) \cite{Blinov2010bk}.
\begin{table}[h]
\caption{\label{Table_1}%
Parameters of the layers.
}
\begin{ruledtabular}
\begin{tabular}{@{}ccc@{}}
Layer & Thickness ($\mu$m) & \begin{tabular}[c]{@{}c@{}}Refractive index\\ at $\lambda = 570$ nm\end{tabular} \\ \midrule
Si$_3$N$_4$ & 0.08 & 2.15 \\
SiO$_2$ & 0.153 & 1.45 \\
PVA & 0.1 & 1.48 \\
5СВ & 10.03 & \begin{tabular}[c]{@{}c@{}}$n_\perp = \sqrt{\varepsilon_\perp} = 1.55$ \\  $n_\parallel = \sqrt{\varepsilon_\parallel} = 1.74$
\end{tabular}
\end{tabular}
\end{ruledtabular}
\end{table}
\begin{figure}[ht]
\includegraphics{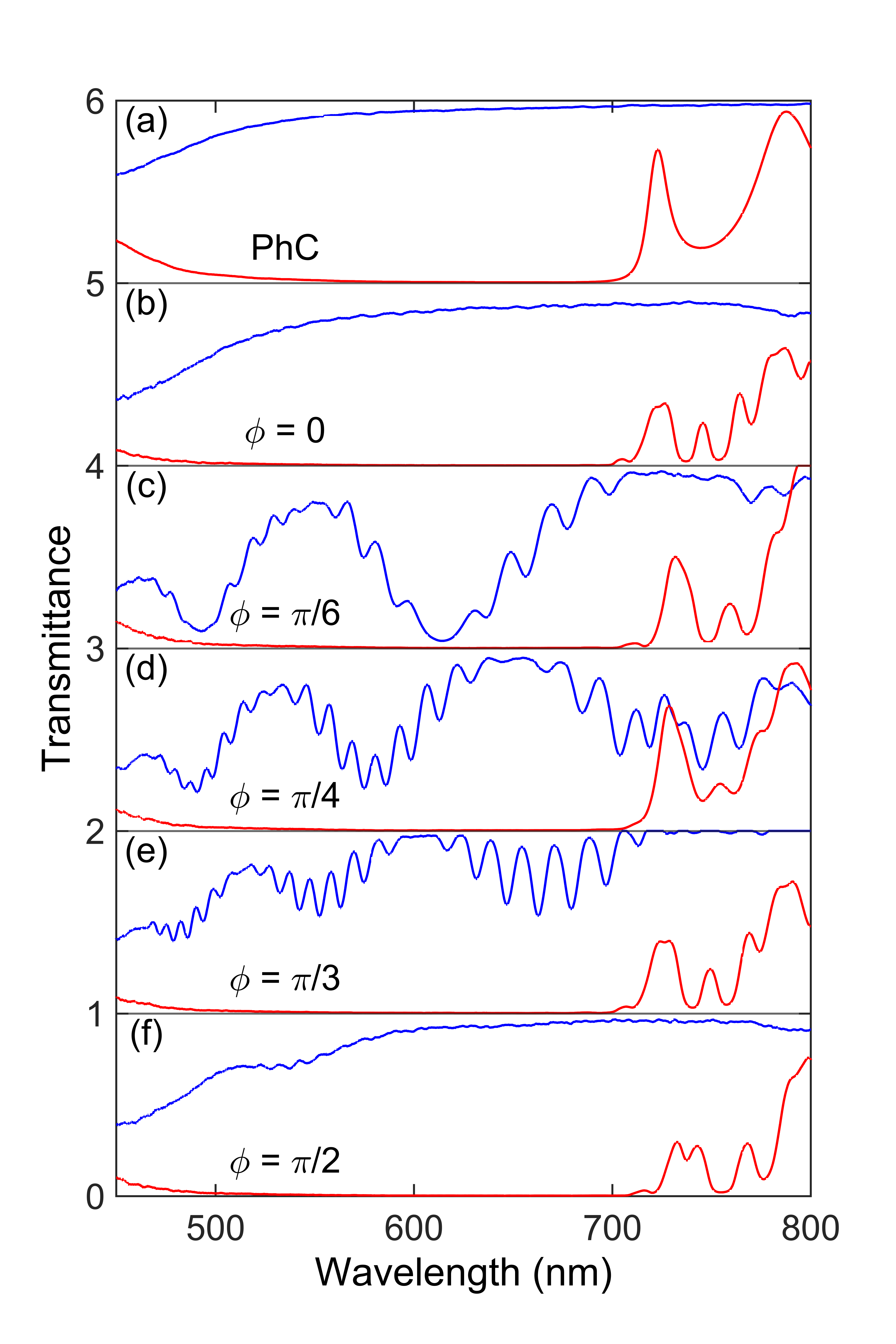}
\caption{ Measured transmittance spectra for the TE (red line) and TM (blue line) waves at the Brewster's angle. The spectra of PhC (a) and microcavity at different values $\phi$ (b-f).}
\label{fig2}
\end{figure}
The polarizing microscopy images of the optical texture of the LC layer confirm the planar LC alignment, see Fig.~\ref{fig1}(c).
The uniform dark texture ensures rubbing directions of PVA layers are parallel to the polarizer or the analyzer, while the maximum intensity of the transmitted light is observed upon rotation of the crossed polarizers by 45$^{\circ}$.
The microcavity is conjugated with hemispherical lenses made of glass with RI $n_{G} = 1.5$. The immersion oil with RI of 1.5 is placed between the glass substrates and lenses to eliminate the air gap.

The experimental set-up for measuring the microcavity transmittance spectra is shown in Fig.~\ref{fig1}(d). 
The incoherent radiation from the source propagates through an optical fiber and a polarizer. After passing through the polarizer, the TE-polarized (TE wave) or TM-polarized (TM wave) radiation is focused on the microcavity. The outgoing radiation is collected in a fiber optic collimator connected to the spectrometer.
The microcavity is heated using the thermostat with temperature controlling by a thermistor. The azimuthal angle $\phi$ of the LC OA orientation is changed by rotation of the sample on the motorized turntable.

\section{Results and Discussion}

\begin{figure*}
\centering\includegraphics[width=\linewidth]{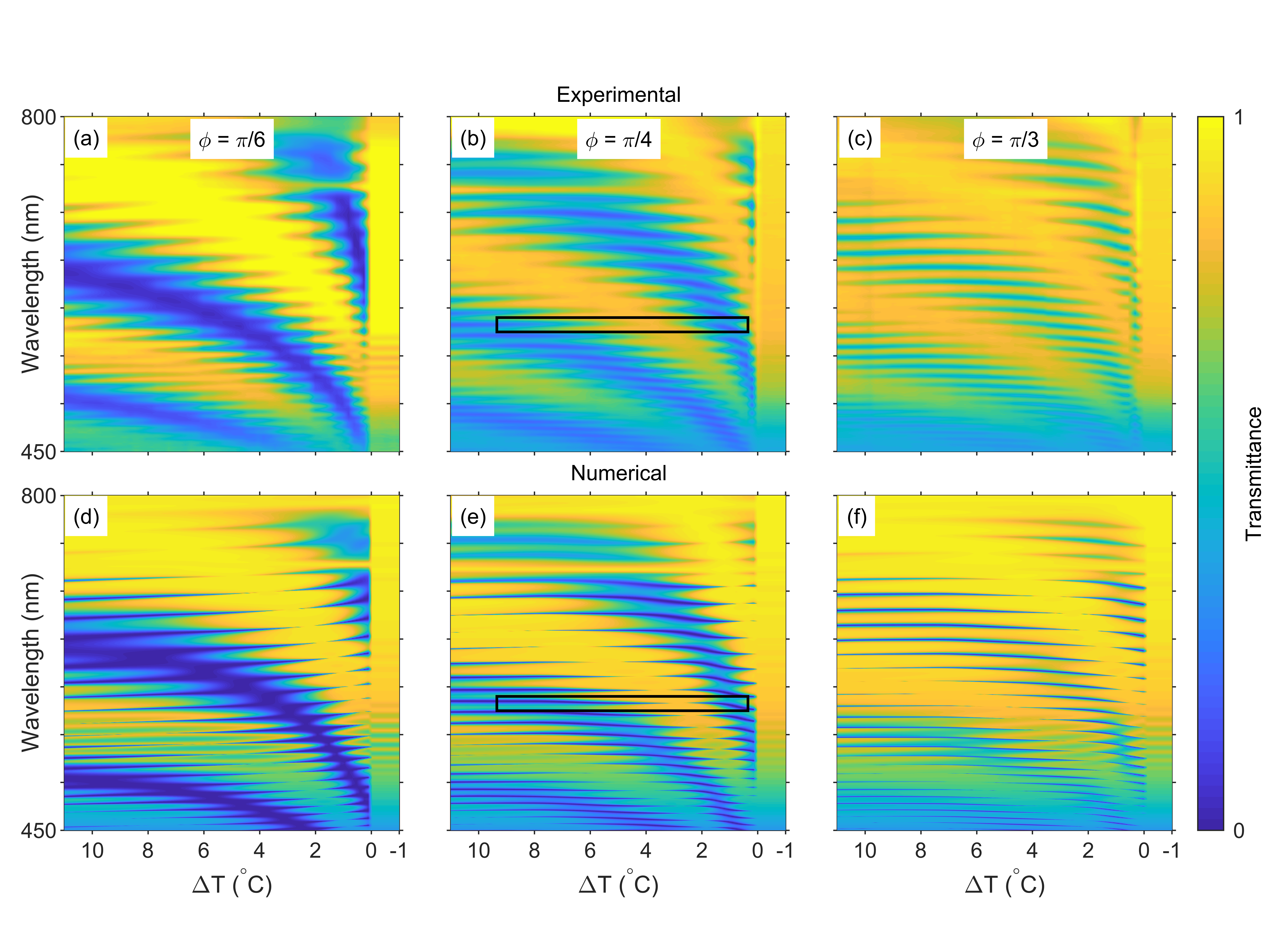}
\caption{ The measured (a-c) and calculated (d-f) dependencies of the microcavity transmittance spectra on the difference $\Delta\text{T} = \text{T}_{\text{c}} -\text{T}$ between the temperature T of the LC layer and the temperature of the LC/isotropic liquid phase transition T$_{\text{c}} \approx 35^{\circ}$C. The black rectangle areas in (b) and (e) are scaled-up in Fig.~\ref{fig4}(a) and Fig.~\ref{fig4}(c), respectively.}
\label{fig3}
\end{figure*}

Figure~\ref{fig2}(a) shows the measured PhC transmittance spectra for TE and TM-waves incident at angle  $\theta_\text{B}~=~\text{asin}{\left[(n_{\text{Si}_3\text{N}_4}/n_{\text{G}}) \sin{(\text{atan}{(n_{\text{SiO}_2}/n_{\text{Si}_3\text{N}_4})})}\right]} \approx 53^{\circ}$.
The wide dip corresponding to the PBG for TE waves is observed, while TM waves pass through the PhC due to the Brewster effect at the Si$_3$N$_4$/SiO$_2$ interfaces \cite{Akhmanov1997bk}. 
The transmittance spectra of the microcavity for different values of the azimuthal angle $\phi$ at room temperature are shown in Figure~\ref{fig2}(b-f).
In the PBG spectral region the multiple resonant dips corresponding to the MC modes are observed when $\phi~\neq~0,~\pi/2$, see Fig.~\ref{fig2}(c-e).
It can be seen that the position and width of the resonant lines are strongly dependent on the value $\phi$.
The width of the resonant line is determined by the total decay rate $\gamma_{tot}~=~\gamma~+~\gamma_{0}$, which is the sum of the radiation decay rate $\gamma$ into the TM polarized continuum and the material loss rate $\gamma_{0}$. 
The cases $\phi~=~0,~\pi/2$ correspond to symmetry-protected BICs with zero radiation decay rate $\gamma = 0$, due to the orthogonality of the localized TE and propagating TM waves \cite{Pankin2020BIC, KrasnovBIC2023}. The resonant dips do not occur in the corresponding spectra, see Fig.~\ref{fig2}(b, f).

The measured temperature transformation of the microcavity spectra for fixed values $\phi~=~\pi/6,~\pi/4,~\pi/3$ is shown in Fig.~\ref{fig3}(a-c). 
It can be seen that position and width of the resonant lines change when temperature $\text{T}$ increases from 24$^{\circ}$C to 35$^{\circ}$C.
The radiation decay rate $\gamma$ is determined by the Poynting vector of TM waves of the resonant mode $\gamma~\propto~|E_x|^2$ at the LC/PVA boundary \cite{Timofeev2018_BIC}.
The value $E_x$ is a sum of the ordinary (o-wave) and extraordinary (e-wave) waves $E_x~=~E_{ox}~+~E_{ex}$.
The polarization vectors $\bm{E_{\text{o,e}}}$ of the o- and e-waves are determined by the direction of the LC OA $\bm{a}$, the permittivities of the o- and e-waves $\varepsilon_{\text{o,e}}(\text{T})$, and the unit vectors in the propagation directions $\bm{\kappa_{\text{o,e}}}(\text{T})~=~[\kappa_{\text{o,ex}}; 0; \kappa_{\text{o,ez}}]$. According to ~\cite{Ignatovich2012} one can write 
\begin{align}
\bm{E_{\text{o}}} & ~=~E_{\text{o}}~\left[\bm{a}~\times\bm{\kappa_{\text{o}}} \right], \nonumber \\ \bm{E_{\text{e}}} & ~=~E_{\text{e}}~\left[ \bm{a}~-~\frac{\varepsilon_{\text{e}}(\alpha)}{\varepsilon_{\text{o}}}\bm{\kappa_{\text{e}}}(\bm{\kappa_{\text{e}}}\bm{a}) \right],
\end{align}
where $\alpha$ is the angle between the vectors $\bm{a}$ and $\bm{\kappa_{\text{e}}}$.
The thermal motion of the LC molecules leads to a change in the permittivities $\varepsilon_o~=~\varepsilon_{\perp}(\text{T})$ and $\varepsilon_{\perp}(\text{T})~\leq~\varepsilon_{\text{e}}(\alpha)~\leq~\varepsilon_{\parallel}(\text{T})$.
For certain values of temperature T the condition $E_x~=~E_{ox}~+~E_{ex}~=~0$ can be satisfied. The radiation decay rate $\gamma$ in this case is equal to zero $\gamma = 0$, and the resonant line collapses. This case corresponds to Friedrich--Wintgen BIC~\cite{Friedrich1985}, also called accidental BIC~\cite{Hsu2013} or parametric BIC~\cite{Koshelev2023}.  
At temperature T$_\text{c}~=~35^{\circ}$C the phase transition of the LC to the isotropic liquid is observed. The TE and TM waves are not mixed, and resonant lines vanish \cite{PankinLPR2021_Temp}.

\begin{figure}[!ht]
\includegraphics{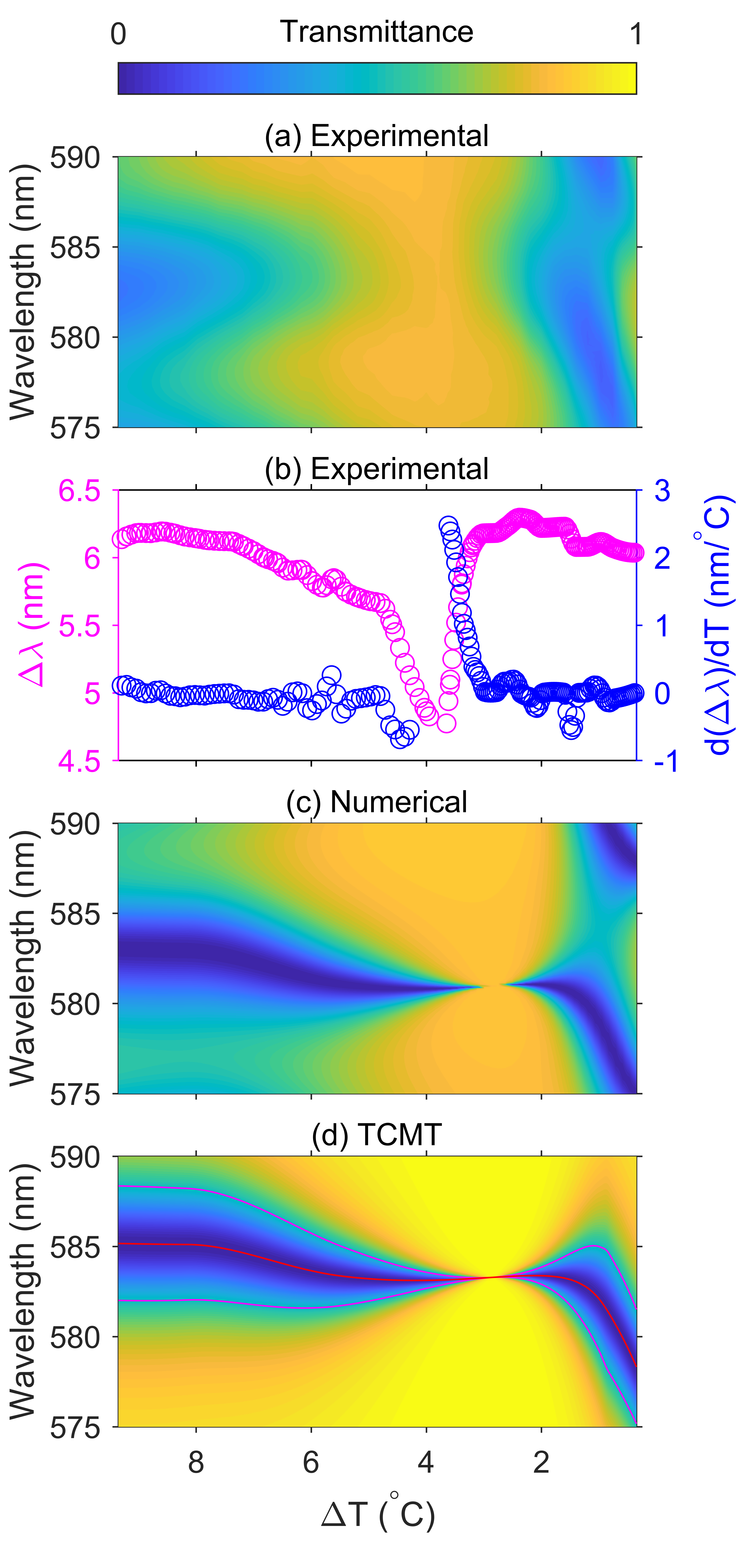}
\caption{ The measured (a), calculated by Berreman's method (c) and using TCMT model (d) dependencies of the microcavity transmittance spectra on the difference $\Delta\text{T} = \text{T}_{\text{c}} -\text{T}$ between the temperature T of the LC layer and the temperature of the LC/isotropic liquid phase transition T$_{\text{c}} \approx 35^{\circ}$C. Subplot~(d) shows $2\pi/\omega_{0}$ (red line); $2\pi/(\omega_{0} \pm \gamma_{tot})
$ (magenta line).  (b) The linewidth $\Delta\lambda$, and its derivative $\text{d}(\Delta\lambda$)/\text{dT} for the resonant line are shown in (a).}
\label{fig4}
\end{figure}

Figure~\ref{fig3}(d-f) presents temperature transformation of the microcavity spectra calculated by the Berreman transfer matrix method \cite{Berreman1972}.
For calculating the spectra the frequency-dependent \cite{luke2015broadband, Gao2013_RI_SIO2, Schnepf2017PVA_RI, tambasov2019structural, wu1993refractive} and temperature-dependent refractive indices \cite{wu1993refractive} were adjusted within 5$\%$ to provide agreement with the experiment. 
In Fig.~\ref{fig4}~(a) we demonstrate the transmittance in the vicinity of the BIC shown in~Fig.~\ref{fig3}~(b) by black rectangle. 
In Fig.~\ref{fig4}~(b) we show the temperature dependence of the resonant linewidth and its temperature derivative.
In Fig.~\ref{fig4}~(b) one can see that at the BIC temperature T the derivative exhibits an abrupt change.
In Fig.~\ref{fig4}~(c) we show the numerical data obtained by the Berreman method in the same range of parameters.

In the framework of the temporal coupled-mode theory (TCMT)~\cite{FanShanhui2003} the LC layer and the PhCs are considered as a resonator and waveguides. The amplitude $a$ of the MC mode obeys the following equations 
\begin{align}\label{a}
    & \frac{da}{dt}=-(i\omega_0+\gamma + \gamma_0)a+\langle d^*|
    \left(
    \begin{array}{c}
         s_1^{\smallp} \\
         {s}_2^{\smallp}
    \end{array}
    \right),
\end{align}

\begin{align}\label{in_out}
    & \left(
\begin{array}{c}
s_{1}^{\smallm} \\
{s}_{2}^{\smallm} \\
\end{array}
\right)
=\widehat{C}
\left(
\begin{array}{c}
s_{1}^{\smallp} \\
{s}_{2}^{\smallp} \\
\end{array}
\right)+ a|d\rangle.
\end{align}
Here $\omega_0$ is the resonant frequency, $|d\rangle$ is the column vector of coupling constants, $s_{1,2}^{\smallpm}$ are amplitudes of TM waves in the PhC waveguides.
In the case of central-plane mirror symmetry, the non-resonant scattering matrix $\widehat{C}$ at the BIC frequency is written as 
\begin{equation}\label{direct}
\widehat{C}=
e^{i\psi}\left(
\begin{array}{cc}
\rho & \pm i\tau\\
\pm i\tau & \rho
\end{array}
\right),
\end{equation}
where $\psi$ and $\rho$ are the phase and amplitude of the complex reflection coefficient, $\tau$ is the  amplitude of the complex transmission coefficient, $\rho^2~+~\tau^2~=~1$.
The coupling constant vector for the case when the MC mode is even with respect to the mirror plane is written as
\begin{equation}
\label{d}
|d\rangle= \left(\begin{array}{l}d_1  \\ d_2 \end{array}\right) =
e^{i\frac{\psi}{2}}
\sqrt{\frac{\gamma }{2(1+\rho)}}
\left(
\begin{array}{l}
\pm\tau-i(1+\rho) \\
\pm\tau- i(1+\rho)
\end{array}
\right), 
\end{equation}
and for the case of odd MC mode it is written as
\begin{equation}
\label{dm}
|d\rangle= \left(\begin{array}{l}d_1  \\ d_2 \end{array}\right) =
e^{i\frac{\psi}{2}}
\sqrt{\frac{\gamma }{2(1+\rho)}}
\left(
\begin{array}{l}
\pm\tau+i(1+\rho) \\
\mp\tau- i(1+\rho)
\end{array}
\right).
\end{equation}
Assuming that harmonic waves $s^{\smallpm} \propto e^{-i\omega t}$ propagate in the waveguides, the Eq.~\eqref{a} and Eq.~\eqref{in_out} yield the final expression for the scattering matrix $\widehat{S}$ in following form 
\begin{equation}\label{S}
\widehat{S}(\text{T},\omega)=\widehat{C}+\frac{|d\rangle\langle d^*|}{i(\omega_0(\text{T}) - \omega) + \gamma(\text{T}) + \gamma_0}.
\end{equation}

The complex eigenfrequency in dependence on temperature $\omega_{r}(\text{T}) = \omega_0(\text{T}) -i\gamma(\text{T})$ was found analytically by solving the eigenvalue problem for an open system, see Supplementary in \cite{Pankin2020BIC}.
The major contribution to the material loss in the fabricated microcavity is due to the conducting transparent layers of aluminum-doped zinc oxide deposited on the glass substrates.  
It cannot be taken into account by solving the eigenvalue problem formulated for the semi-infinite PhCs, therefore the rate of material loss $\gamma_{0}$ is fitted to be consistent with the numerical spectra. 
In Fig.~\ref{fig4}~(d) we demonstrate the TCMT solution with fitted $\gamma_0$. One can see a good agreement between Fig.~\ref{fig4}~(d) and  Fig.~\ref{fig4}~(c).
Although the presence of the conducting layers decrease the maximum possible quality factor $Q = \omega_0/2\gamma_{tot}$ to the value $Q_{max} = \omega_0/2\gamma_0$ at the BIC frequency, it ensures the voltage-tunable Q factor which has been demonstrated in \cite{KrasnovBIC2023}.

\section{conclusion}

In this work we demonstrated an optical bound state in the continuum in a photonic crystal microcavity tunable by heating the anisotropic liquid crystal  resonant layer. We experimentally measured the temperature dependencies of the transmittance spectrum at Brewster's angle and observed multiple vanishing resonant lines which indicate the occurrence of optical bound states in the continuum.
The obtained dependencies are explained theoretically with application of the temporal coupled mode theory and rigorous Berreman's transfer matrix method.
It is found that in the point of optical bound state in the continuum the resonant linewidth has an abrupt change which can be employed for engineering temperature sensors. 
 
Acknowledgments. This work was supported by the Russian Science Foundation under grant no 22-22-00687.
The authors would like to express their special gratitude to the Krasnoyarsk Regional Center for Collective Use of the Federal Research Center “Krasnoyarsk Scientific Center, Siberian Branch of the Russian Academy of Sciences” for providing equipment within this project.

Disclosures. The authors declare no conflicts of interest.

Data Availability Statement. The data that support the findings of this study are available from the corresponding author, P.S.P., upon reasonable request.

\nocite{}

\bibliography{Manuscript.bbl}

\begin{thebibliography}{54}%
\makeatletter
\providecommand \@ifxundefined [1]{%
 \@ifx{#1\undefined}
}%
\providecommand \@ifnum [1]{%
 \ifnum #1\expandafter \@firstoftwo
 \else \expandafter \@secondoftwo
 \fi
}%
\providecommand \@ifx [1]{%
 \ifx #1\expandafter \@firstoftwo
 \else \expandafter \@secondoftwo
 \fi
}%
\providecommand \natexlab [1]{#1}%
\providecommand \enquote  [1]{``#1''}%
\providecommand \bibnamefont  [1]{#1}%
\providecommand \bibfnamefont [1]{#1}%
\providecommand \citenamefont [1]{#1}%
\providecommand \href@noop [0]{\@secondoftwo}%
\providecommand \href [0]{\begingroup \@sanitize@url \@href}%
\providecommand \@href[1]{\@@startlink{#1}\@@href}%
\providecommand \@@href[1]{\endgroup#1\@@endlink}%
\providecommand \@sanitize@url [0]{\catcode `\\12\catcode `\$12\catcode
  `\&12\catcode `\#12\catcode `\^12\catcode `\_12\catcode `\%12\relax}%
\providecommand \@@startlink[1]{}%
\providecommand \@@endlink[0]{}%
\providecommand \url  [0]{\begingroup\@sanitize@url \@url }%
\providecommand \@url [1]{\endgroup\@href {#1}{\urlprefix }}%
\providecommand \urlprefix  [0]{URL }%
\providecommand \Eprint [0]{\href }%
\providecommand \doibase [0]{https://doi.org/}%
\providecommand \selectlanguage [0]{\@gobble}%
\providecommand \bibinfo  [0]{\@secondoftwo}%
\providecommand \bibfield  [0]{\@secondoftwo}%
\providecommand \translation [1]{[#1]}%
\providecommand \BibitemOpen [0]{}%
\providecommand \bibitemStop [0]{}%
\providecommand \bibitemNoStop [0]{.\EOS\space}%
\providecommand \EOS [0]{\spacefactor3000\relax}%
\providecommand \BibitemShut  [1]{\csname bibitem#1\endcsname}%
\let\auto@bib@innerbib\@empty
\bibitem [{\citenamefont {Joannopoulos}\ \emph {et~al.}(2008)\citenamefont
  {Joannopoulos}, \citenamefont {Johnson}, \citenamefont {Winn},\ and\
  \citenamefont {Meade}}]{Joannopoulos2008bk}%
  \BibitemOpen
  \bibfield  {author} {\bibinfo {author} {\bibfnamefont {J.~D.}\ \bibnamefont
  {Joannopoulos}}, \bibinfo {author} {\bibfnamefont {S.~G.}\ \bibnamefont
  {Johnson}}, \bibinfo {author} {\bibfnamefont {J.~N.}\ \bibnamefont {Winn}},\
  and\ \bibinfo {author} {\bibfnamefont {R.~D.}\ \bibnamefont {Meade}},\
  }\href@noop {} {\emph {\bibinfo {title} {{Photonic Crystals: Molding the Flow
  of Light (Second Edition)}}}}\ (\bibinfo  {publisher} {Princeton University
  Press},\ \bibinfo {address} {Princeton, NJ, USA},\ \bibinfo {year} {2008})\
  p.\ \bibinfo {pages} {304}\BibitemShut {NoStop}%
\bibitem [{\citenamefont {Kavokin}\ \emph {et~al.}(2017)\citenamefont
  {Kavokin}, \citenamefont {Baumberg}, \citenamefont {Malpuech},\ and\
  \citenamefont {Laussy}}]{kavokin2017microcavities}%
  \BibitemOpen
  \bibfield  {author} {\bibinfo {author} {\bibfnamefont {A.~V.}\ \bibnamefont
  {Kavokin}}, \bibinfo {author} {\bibfnamefont {J.~J.}\ \bibnamefont
  {Baumberg}}, \bibinfo {author} {\bibfnamefont {G.}~\bibnamefont {Malpuech}},\
  and\ \bibinfo {author} {\bibfnamefont {F.~P.}\ \bibnamefont {Laussy}},\
  }\href@noop {} {\emph {\bibinfo {title} {{Microcavities}}}},\ Vol.~\bibinfo
  {volume} {21}\ (\bibinfo  {publisher} {Oxford university press},\ \bibinfo
  {year} {2017})\BibitemShut {NoStop}%
\bibitem [{\citenamefont {Timofeev}\ \emph {et~al.}(2018)\citenamefont
  {Timofeev}, \citenamefont {Maksimov},\ and\ \citenamefont
  {Sadreev}}]{Timofeev2018_BIC}%
  \BibitemOpen
  \bibfield  {author} {\bibinfo {author} {\bibfnamefont {I.~V.}\ \bibnamefont
  {Timofeev}}, \bibinfo {author} {\bibfnamefont {D.~N.}\ \bibnamefont
  {Maksimov}},\ and\ \bibinfo {author} {\bibfnamefont {A.~F.}\ \bibnamefont
  {Sadreev}},\ }\bibfield  {title} {\bibinfo {title} {{Optical defect mode with
  tunable Q factor in a one-dimensional anisotropic photonic crystal}},\
  }\href@noop {} {\bibfield  {journal} {\bibinfo  {journal} {Physical Review
  B}\ }\textbf {\bibinfo {volume} {97}},\ \bibinfo {pages} {24306} (\bibinfo
  {year} {2018})}\BibitemShut {NoStop}%
\bibitem [{\citenamefont {Pankin}\ \emph
  {et~al.}(2020{\natexlab{a}})\citenamefont {Pankin}, \citenamefont {Wu},
  \citenamefont {Yang}, \citenamefont {Chen}, \citenamefont {Timofeev},\ and\
  \citenamefont {Sadreev}}]{Pankin2020BIC}%
  \BibitemOpen
  \bibfield  {author} {\bibinfo {author} {\bibfnamefont {P.~S.}\ \bibnamefont
  {Pankin}}, \bibinfo {author} {\bibfnamefont {B.-R.}\ \bibnamefont {Wu}},
  \bibinfo {author} {\bibfnamefont {J.-H.}\ \bibnamefont {Yang}}, \bibinfo
  {author} {\bibfnamefont {K.-P.}\ \bibnamefont {Chen}}, \bibinfo {author}
  {\bibfnamefont {I.~V.}\ \bibnamefont {Timofeev}},\ and\ \bibinfo {author}
  {\bibfnamefont {A.~F.}\ \bibnamefont {Sadreev}},\ }\bibfield  {title}
  {\bibinfo {title} {{One-dimensional photonic bound states in the
  continuum}},\ }\href {https://doi.org/10.1038/s42005-020-0353-z} {\bibfield
  {journal} {\bibinfo  {journal} {Communications Physics}\ }\textbf {\bibinfo
  {volume} {3}},\ \bibinfo {pages} {91} (\bibinfo {year}
  {2020}{\natexlab{a}})}\BibitemShut {NoStop}%
\bibitem [{\citenamefont {Pankin}\ \emph
  {et~al.}(2020{\natexlab{b}})\citenamefont {Pankin}, \citenamefont {Maksimov},
  \citenamefont {Chen},\ and\ \citenamefont {Timofeev}}]{Pankin2020Fano}%
  \BibitemOpen
  \bibfield  {author} {\bibinfo {author} {\bibfnamefont {P.~S.}\ \bibnamefont
  {Pankin}}, \bibinfo {author} {\bibfnamefont {D.~N.}\ \bibnamefont
  {Maksimov}}, \bibinfo {author} {\bibfnamefont {K.~P.}\ \bibnamefont {Chen}},\
  and\ \bibinfo {author} {\bibfnamefont {I.~V.}\ \bibnamefont {Timofeev}},\
  }\bibfield  {title} {\bibinfo {title} {{Fano feature induced by a bound state
  in the continuum via resonant state expansion}},\ }\href
  {https://doi.org/10.1038/s41598-020-70654-2} {\bibfield  {journal} {\bibinfo
  {journal} {Scientific Reports}\ }\textbf {\bibinfo {volume} {10}},\ \bibinfo
  {pages} {13691} (\bibinfo {year} {2020}{\natexlab{b}})}\BibitemShut {NoStop}%
\bibitem [{\citenamefont {Ignatyeva}\ and\ \citenamefont
  {Belotelov}(2020)}]{Ignatyeva2020BIC}%
  \BibitemOpen
  \bibfield  {author} {\bibinfo {author} {\bibfnamefont {D.~O.}\ \bibnamefont
  {Ignatyeva}}\ and\ \bibinfo {author} {\bibfnamefont {V.~I.}\ \bibnamefont
  {Belotelov}},\ }\bibfield  {title} {\bibinfo {title} {{Bound states in the
  continuum enable modulation of light intensity in the Faraday
  configuration}},\ }\href {https://doi.org/10.1364/OL.404159} {\bibfield
  {journal} {\bibinfo  {journal} {Optics Letters}\ }\textbf {\bibinfo {volume}
  {45}},\ \bibinfo {pages} {6422} (\bibinfo {year} {2020})}\BibitemShut
  {NoStop}%
\bibitem [{\citenamefont {Wu}\ \emph {et~al.}(2021)\citenamefont {Wu},
  \citenamefont {Yang}, \citenamefont {Pankin}, \citenamefont {Huang},
  \citenamefont {Lee}, \citenamefont {Maksimov}, \citenamefont {Timofeev},\
  and\ \citenamefont {Chen}}]{PankinLPR2021_Temp}%
  \BibitemOpen
  \bibfield  {author} {\bibinfo {author} {\bibfnamefont {B.}~\bibnamefont
  {Wu}}, \bibinfo {author} {\bibfnamefont {J.}~\bibnamefont {Yang}}, \bibinfo
  {author} {\bibfnamefont {P.~S.}\ \bibnamefont {Pankin}}, \bibinfo {author}
  {\bibfnamefont {C.}~\bibnamefont {Huang}}, \bibinfo {author} {\bibfnamefont
  {W.}~\bibnamefont {Lee}}, \bibinfo {author} {\bibfnamefont {D.~N.}\
  \bibnamefont {Maksimov}}, \bibinfo {author} {\bibfnamefont {I.~V.}\
  \bibnamefont {Timofeev}},\ and\ \bibinfo {author} {\bibfnamefont
  {K.}~\bibnamefont {Chen}},\ }\bibfield  {title} {\bibinfo {title}
  {{Quasi‐Bound States in the Continuum with Temperature‐Tunable Q Factors
  and Critical Coupling Point at Brewster's Angle}},\ }\href
  {https://doi.org/10.1002/lpor.202000290} {\bibfield  {journal} {\bibinfo
  {journal} {Laser and Photonics Reviews}\ }\textbf {\bibinfo {volume} {15}},\
  \bibinfo {pages} {2000290} (\bibinfo {year} {2021})}\BibitemShut {NoStop}%
\bibitem [{\citenamefont {Pankin}\ \emph {et~al.}(2022)\citenamefont {Pankin},
  \citenamefont {Maksimov},\ and\ \citenamefont {Timofeev}}]{Pankin_JOSAB2022}%
  \BibitemOpen
  \bibfield  {author} {\bibinfo {author} {\bibfnamefont {P.~S.}\ \bibnamefont
  {Pankin}}, \bibinfo {author} {\bibfnamefont {D.~N.}\ \bibnamefont
  {Maksimov}},\ and\ \bibinfo {author} {\bibfnamefont {I.~V.}\ \bibnamefont
  {Timofeev}},\ }\bibfield  {title} {\bibinfo {title} {{Bound state in the
  continuum in an anisotropic photonic crystal supported by a full-wave phase
  plate}},\ }\href {https://doi.org/10.1364/JOSAB.451034} {\bibfield  {journal}
  {\bibinfo  {journal} {Journal of the Optical Society of America B}\ }\textbf
  {\bibinfo {volume} {39}},\ \bibinfo {pages} {968} (\bibinfo {year}
  {2022})}\BibitemShut {NoStop}%
\bibitem [{\citenamefont {Nabol}\ \emph {et~al.}(2022)\citenamefont {Nabol},
  \citenamefont {Pankin}, \citenamefont {Maksimov},\ and\ \citenamefont
  {Timofeev}}]{nabol2022fabryperot}%
  \BibitemOpen
  \bibfield  {author} {\bibinfo {author} {\bibfnamefont {S.~V.}\ \bibnamefont
  {Nabol}}, \bibinfo {author} {\bibfnamefont {P.~S.}\ \bibnamefont {Pankin}},
  \bibinfo {author} {\bibfnamefont {D.~N.}\ \bibnamefont {Maksimov}},\ and\
  \bibinfo {author} {\bibfnamefont {I.~V.}\ \bibnamefont {Timofeev}},\
  }\bibfield  {title} {\bibinfo {title} {Fabry-perot bound states in the
  continuum in an anisotropic photonic crystal},\ }\href@noop {} {\bibfield
  {journal} {\bibinfo  {journal} {Physical Review B}\ }\textbf {\bibinfo
  {volume} {106}},\ \bibinfo {pages} {245403} (\bibinfo {year}
  {2022})}\BibitemShut {NoStop}%
\bibitem [{\citenamefont {Krasnov}\ \emph {et~al.}(2023)\citenamefont
  {Krasnov}, \citenamefont {Pankin}, \citenamefont {Buzin}, \citenamefont
  {Romanenko}, \citenamefont {Sutormin}, \citenamefont {Zelenov}, \citenamefont
  {Masyugin}, \citenamefont {Volochaev}, \citenamefont {Vetrov},\ and\
  \citenamefont {Timofeev}}]{KrasnovBIC2023}%
  \BibitemOpen
  \bibfield  {author} {\bibinfo {author} {\bibfnamefont {A.~I.}\ \bibnamefont
  {Krasnov}}, \bibinfo {author} {\bibfnamefont {P.~S.}\ \bibnamefont {Pankin}},
  \bibinfo {author} {\bibfnamefont {D.~S.}\ \bibnamefont {Buzin}}, \bibinfo
  {author} {\bibfnamefont {G.~A.}\ \bibnamefont {Romanenko}}, \bibinfo {author}
  {\bibfnamefont {V.~S.}\ \bibnamefont {Sutormin}}, \bibinfo {author}
  {\bibfnamefont {F.~V.}\ \bibnamefont {Zelenov}}, \bibinfo {author}
  {\bibfnamefont {A.~N.}\ \bibnamefont {Masyugin}}, \bibinfo {author}
  {\bibfnamefont {M.~N.}\ \bibnamefont {Volochaev}}, \bibinfo {author}
  {\bibfnamefont {S.~Y.}\ \bibnamefont {Vetrov}},\ and\ \bibinfo {author}
  {\bibfnamefont {I.~V.}\ \bibnamefont {Timofeev}},\ }\bibfield  {title}
  {\bibinfo {title} {{Voltage-tunable Q factor in a photonic crystal
  microcavity}},\ }\href {https://doi.org/10.1364/OL.479431} {\bibfield
  {journal} {\bibinfo  {journal} {Optics Letters}\ }\textbf {\bibinfo {volume}
  {48}},\ \bibinfo {pages} {1666} (\bibinfo {year} {2023})}\BibitemShut
  {NoStop}%
\bibitem [{\citenamefont {Hsu}\ \emph {et~al.}(2016)\citenamefont {Hsu},
  \citenamefont {Zhen}, \citenamefont {Stone}, \citenamefont {Joannopoulos},\
  and\ \citenamefont {Solja{\v{c}}i{\'{c}}}}]{HsuChiaWei2016}%
  \BibitemOpen
  \bibfield  {author} {\bibinfo {author} {\bibfnamefont {C.~W.}\ \bibnamefont
  {Hsu}}, \bibinfo {author} {\bibfnamefont {B.}~\bibnamefont {Zhen}}, \bibinfo
  {author} {\bibfnamefont {A.~D.}\ \bibnamefont {Stone}}, \bibinfo {author}
  {\bibfnamefont {J.~D.}\ \bibnamefont {Joannopoulos}},\ and\ \bibinfo {author}
  {\bibfnamefont {M.}~\bibnamefont {Solja{\v{c}}i{\'{c}}}},\ }\bibfield
  {title} {\bibinfo {title} {{Bound states in the continuum}},\ }\href
  {https://doi.org/10.1038/natrevmats.2016.48} {\bibfield  {journal} {\bibinfo
  {journal} {Nat. Rev. Mater.}\ }\textbf {\bibinfo {volume} {1}},\ \bibinfo
  {pages} {16048} (\bibinfo {year} {2016})}\BibitemShut {NoStop}%
\bibitem [{\citenamefont {Sadreev}(2021)}]{sadreev2021interference}%
  \BibitemOpen
  \bibfield  {author} {\bibinfo {author} {\bibfnamefont {A.~F.}\ \bibnamefont
  {Sadreev}},\ }\bibfield  {title} {\bibinfo {title} {{Interference traps waves
  in open system: Bound states in the continuum}},\ }\href@noop {} {\bibfield
  {journal} {\bibinfo  {journal} {Reports on Progress in Physics}\ } (\bibinfo
  {year} {2021})}\BibitemShut {NoStop}%
\bibitem [{\citenamefont {Koshelev}\ \emph {et~al.}(2023)\citenamefont
  {Koshelev}, \citenamefont {Sadrieva}, \citenamefont {Shcherbakov},
  \citenamefont {Kivshar},\ and\ \citenamefont {Bogdanov}}]{Koshelev2023}%
  \BibitemOpen
  \bibfield  {author} {\bibinfo {author} {\bibfnamefont {K.~L.}\ \bibnamefont
  {Koshelev}}, \bibinfo {author} {\bibfnamefont {Z.~F.}\ \bibnamefont
  {Sadrieva}}, \bibinfo {author} {\bibfnamefont {A.~A.}\ \bibnamefont
  {Shcherbakov}}, \bibinfo {author} {\bibfnamefont {Y.}~\bibnamefont
  {Kivshar}},\ and\ \bibinfo {author} {\bibfnamefont {A.~A.}\ \bibnamefont
  {Bogdanov}},\ }\bibfield  {title} {\bibinfo {title} {Bound states in the
  continuum in photonic structures},\ }\bibfield  {journal} {\bibinfo
  {journal} {Physics-Uspekhi}\ }\textbf {\bibinfo {volume} {66}},\ \href
  {https://doi.org/10.3367/UFNe.2021.12.039120} {10.3367/UFNe.2021.12.039120}
  (\bibinfo {year} {2023})\BibitemShut {NoStop}%
\bibitem [{\citenamefont {Azzam}\ and\ \citenamefont
  {Kildishev}(2021)}]{Azzam2020BICReview}%
  \BibitemOpen
  \bibfield  {author} {\bibinfo {author} {\bibfnamefont {S.~I.}\ \bibnamefont
  {Azzam}}\ and\ \bibinfo {author} {\bibfnamefont {A.~V.}\ \bibnamefont
  {Kildishev}},\ }\bibfield  {title} {\bibinfo {title} {{Photonic Bound States
  in the Continuum: From Basics to Applications}},\ }\href
  {https://doi.org/10.1002/adom.202001469} {\bibfield  {journal} {\bibinfo
  {journal} {Advanced Optical Materials}\ }\textbf {\bibinfo {volume} {9}},\
  \bibinfo {pages} {2001469} (\bibinfo {year} {2021})}\BibitemShut {NoStop}%
\bibitem [{\citenamefont {Joseph}\ \emph {et~al.}(2021)\citenamefont {Joseph},
  \citenamefont {Pandey}, \citenamefont {Sarkar},\ and\ \citenamefont
  {Joseph}}]{Joseph2021}%
  \BibitemOpen
  \bibfield  {author} {\bibinfo {author} {\bibfnamefont {S.}~\bibnamefont
  {Joseph}}, \bibinfo {author} {\bibfnamefont {S.}~\bibnamefont {Pandey}},
  \bibinfo {author} {\bibfnamefont {S.}~\bibnamefont {Sarkar}},\ and\ \bibinfo
  {author} {\bibfnamefont {J.}~\bibnamefont {Joseph}},\ }\bibfield  {title}
  {\bibinfo {title} {{Bound states in the continuum in resonant nanostructures:
  an overview of engineered materials for tailored applications}},\ }\href
  {https://doi.org/10.1515/nanoph-2021-0387} {\bibfield  {journal} {\bibinfo
  {journal} {Nanophotonics}\ }\textbf {\bibinfo {volume} {10}},\ \bibinfo
  {pages} {4175} (\bibinfo {year} {2021})}\BibitemShut {NoStop}%
\bibitem [{\citenamefont {Kodigala}\ \emph {et~al.}(2017)\citenamefont
  {Kodigala}, \citenamefont {Lepetit}, \citenamefont {Gu}, \citenamefont
  {Bahari}, \citenamefont {Fainman},\ and\ \citenamefont
  {Kant{\'{e}}}}]{Kodigala17}%
  \BibitemOpen
  \bibfield  {author} {\bibinfo {author} {\bibfnamefont {A.}~\bibnamefont
  {Kodigala}}, \bibinfo {author} {\bibfnamefont {T.}~\bibnamefont {Lepetit}},
  \bibinfo {author} {\bibfnamefont {Q.}~\bibnamefont {Gu}}, \bibinfo {author}
  {\bibfnamefont {B.}~\bibnamefont {Bahari}}, \bibinfo {author} {\bibfnamefont
  {Y.}~\bibnamefont {Fainman}},\ and\ \bibinfo {author} {\bibfnamefont
  {B.}~\bibnamefont {Kant{\'{e}}}},\ }\bibfield  {title} {\bibinfo {title}
  {{Lasing action from photonic bound states in continuum}},\ }\href
  {https://doi.org/10.1038/nature20799} {\bibfield  {journal} {\bibinfo
  {journal} {Nature}\ }\textbf {\bibinfo {volume} {541}},\ \bibinfo {pages}
  {196} (\bibinfo {year} {2017})}\BibitemShut {NoStop}%
\bibitem [{\citenamefont {Huang}\ \emph
  {et~al.}(2020{\natexlab{a}})\citenamefont {Huang}, \citenamefont {Zhang},
  \citenamefont {Xiao}, \citenamefont {Wang}, \citenamefont {Fan},
  \citenamefont {Liu}, \citenamefont {Zhang}, \citenamefont {Qu}, \citenamefont
  {Ji}, \citenamefont {Han}, \citenamefont {Ge}, \citenamefont {Kivshar},\ and\
  \citenamefont {Song}}]{Kivshar_2020_Science_BIC_Laser}%
  \BibitemOpen
  \bibfield  {author} {\bibinfo {author} {\bibfnamefont {C.}~\bibnamefont
  {Huang}}, \bibinfo {author} {\bibfnamefont {C.}~\bibnamefont {Zhang}},
  \bibinfo {author} {\bibfnamefont {S.}~\bibnamefont {Xiao}}, \bibinfo {author}
  {\bibfnamefont {Y.}~\bibnamefont {Wang}}, \bibinfo {author} {\bibfnamefont
  {Y.}~\bibnamefont {Fan}}, \bibinfo {author} {\bibfnamefont {Y.}~\bibnamefont
  {Liu}}, \bibinfo {author} {\bibfnamefont {N.}~\bibnamefont {Zhang}}, \bibinfo
  {author} {\bibfnamefont {G.}~\bibnamefont {Qu}}, \bibinfo {author}
  {\bibfnamefont {H.}~\bibnamefont {Ji}}, \bibinfo {author} {\bibfnamefont
  {J.}~\bibnamefont {Han}}, \bibinfo {author} {\bibfnamefont {L.}~\bibnamefont
  {Ge}}, \bibinfo {author} {\bibfnamefont {Y.}~\bibnamefont {Kivshar}},\ and\
  \bibinfo {author} {\bibfnamefont {Q.}~\bibnamefont {Song}},\ }\bibfield
  {title} {\bibinfo {title} {{Ultrafast control of vortex microlasers}},\
  }\href {https://doi.org/10.1126/science.aba4597} {\bibfield  {journal}
  {\bibinfo  {journal} {Science}\ }\textbf {\bibinfo {volume} {367}},\ \bibinfo
  {pages} {1018} (\bibinfo {year} {2020}{\natexlab{a}})}\BibitemShut {NoStop}%
\bibitem [{\citenamefont {Yang}\ \emph {et~al.}(2021)\citenamefont {Yang},
  \citenamefont {Huang}, \citenamefont {Maksimov}, \citenamefont {Pankin},
  \citenamefont {Timofeev}, \citenamefont {Hong}, \citenamefont {Li},
  \citenamefont {Chen}, \citenamefont {Hsu}, \citenamefont {Liu}, \citenamefont
  {Lu}, \citenamefont {Lin}, \citenamefont {Yang},\ and\ \citenamefont
  {Chen}}]{Pankin_LPR2021}%
  \BibitemOpen
  \bibfield  {author} {\bibinfo {author} {\bibfnamefont {J.}~\bibnamefont
  {Yang}}, \bibinfo {author} {\bibfnamefont {Z.}~\bibnamefont {Huang}},
  \bibinfo {author} {\bibfnamefont {D.~N.}\ \bibnamefont {Maksimov}}, \bibinfo
  {author} {\bibfnamefont {P.~S.}\ \bibnamefont {Pankin}}, \bibinfo {author}
  {\bibfnamefont {I.~V.}\ \bibnamefont {Timofeev}}, \bibinfo {author}
  {\bibfnamefont {K.}~\bibnamefont {Hong}}, \bibinfo {author} {\bibfnamefont
  {H.}~\bibnamefont {Li}}, \bibinfo {author} {\bibfnamefont {J.}~\bibnamefont
  {Chen}}, \bibinfo {author} {\bibfnamefont {C.}~\bibnamefont {Hsu}}, \bibinfo
  {author} {\bibfnamefont {Y.}~\bibnamefont {Liu}}, \bibinfo {author}
  {\bibfnamefont {T.}~\bibnamefont {Lu}}, \bibinfo {author} {\bibfnamefont
  {T.}~\bibnamefont {Lin}}, \bibinfo {author} {\bibfnamefont {C.}~\bibnamefont
  {Yang}},\ and\ \bibinfo {author} {\bibfnamefont {K.}~\bibnamefont {Chen}},\
  }\bibfield  {title} {\bibinfo {title} {{Low‐Threshold Bound State in the
  Continuum Lasers in Hybrid Lattice Resonance Metasurfaces}},\ }\href
  {https://doi.org/10.1002/lpor.202100118} {\bibfield  {journal} {\bibinfo
  {journal} {Laser and Photonics Reviews}\ }\textbf {\bibinfo {volume} {15}},\
  \bibinfo {pages} {2100118} (\bibinfo {year} {2021})}\BibitemShut {NoStop}%
\bibitem [{\citenamefont {Huang}\ \emph
  {et~al.}(2020{\natexlab{b}})\citenamefont {Huang}, \citenamefont {Zhang},
  \citenamefont {Xiao}, \citenamefont {Wang}, \citenamefont {Fan},
  \citenamefont {Liu}, \citenamefont {Zhang}, \citenamefont {Qu}, \citenamefont
  {Ji}, \citenamefont {Han} \emph {et~al.}}]{huang2020vortex}%
  \BibitemOpen
  \bibfield  {author} {\bibinfo {author} {\bibfnamefont {C.}~\bibnamefont
  {Huang}}, \bibinfo {author} {\bibfnamefont {C.}~\bibnamefont {Zhang}},
  \bibinfo {author} {\bibfnamefont {S.}~\bibnamefont {Xiao}}, \bibinfo {author}
  {\bibfnamefont {Y.}~\bibnamefont {Wang}}, \bibinfo {author} {\bibfnamefont
  {Y.}~\bibnamefont {Fan}}, \bibinfo {author} {\bibfnamefont {Y.}~\bibnamefont
  {Liu}}, \bibinfo {author} {\bibfnamefont {N.}~\bibnamefont {Zhang}}, \bibinfo
  {author} {\bibfnamefont {G.}~\bibnamefont {Qu}}, \bibinfo {author}
  {\bibfnamefont {H.}~\bibnamefont {Ji}}, \bibinfo {author} {\bibfnamefont
  {J.}~\bibnamefont {Han}}, \emph {et~al.},\ }\bibfield  {title} {\bibinfo
  {title} {Ultrafast control of vortex microlasers},\ }\href@noop {} {\bibfield
   {journal} {\bibinfo  {journal} {Science}\ }\textbf {\bibinfo {volume}
  {367}},\ \bibinfo {pages} {1018} (\bibinfo {year}
  {2020}{\natexlab{b}})}\BibitemShut {NoStop}%
\bibitem [{\citenamefont {Hsu}\ \emph {et~al.}(2013)\citenamefont {Hsu},
  \citenamefont {Zhen}, \citenamefont {Lee}, \citenamefont {Chua},
  \citenamefont {Johnson}, \citenamefont {Joannopoulos},\ and\ \citenamefont
  {Solja{\v{c}}i{\'{c}}}}]{Hsu2013}%
  \BibitemOpen
  \bibfield  {author} {\bibinfo {author} {\bibfnamefont {C.~W.}\ \bibnamefont
  {Hsu}}, \bibinfo {author} {\bibfnamefont {B.}~\bibnamefont {Zhen}}, \bibinfo
  {author} {\bibfnamefont {J.}~\bibnamefont {Lee}}, \bibinfo {author}
  {\bibfnamefont {S.~L.}\ \bibnamefont {Chua}}, \bibinfo {author}
  {\bibfnamefont {S.~G.}\ \bibnamefont {Johnson}}, \bibinfo {author}
  {\bibfnamefont {J.~D.}\ \bibnamefont {Joannopoulos}},\ and\ \bibinfo {author}
  {\bibfnamefont {M.}~\bibnamefont {Solja{\v{c}}i{\'{c}}}},\ }\bibfield
  {title} {\bibinfo {title} {{Observation of trapped light within the radiation
  continuum.}},\ }\href {https://doi.org/10.1038/nature12289} {\bibfield
  {journal} {\bibinfo  {journal} {Nature}\ }\textbf {\bibinfo {volume} {499}},\
  \bibinfo {pages} {188} (\bibinfo {year} {2013})}\BibitemShut {NoStop}%
\bibitem [{\citenamefont {Bezus}\ \emph {et~al.}(2018)\citenamefont {Bezus},
  \citenamefont {Bykov},\ and\ \citenamefont {Doskolovich}}]{Bezus2018_BIC}%
  \BibitemOpen
  \bibfield  {author} {\bibinfo {author} {\bibfnamefont {E.~A.}\ \bibnamefont
  {Bezus}}, \bibinfo {author} {\bibfnamefont {D.~A.}\ \bibnamefont {Bykov}},\
  and\ \bibinfo {author} {\bibfnamefont {L.~L.}\ \bibnamefont {Doskolovich}},\
  }\bibfield  {title} {\bibinfo {title} {{Bound states in the continuum and
  high-Q resonances supported by a dielectric ridge on a slab waveguide}},\
  }\href@noop {} {\bibfield  {journal} {\bibinfo  {journal} {Photonics
  Research}\ }\textbf {\bibinfo {volume} {6}},\ \bibinfo {pages} {1084}
  (\bibinfo {year} {2018})}\BibitemShut {NoStop}%
\bibitem [{\citenamefont {Gomis-Bresco}\ \emph {et~al.}(2017)\citenamefont
  {Gomis-Bresco}, \citenamefont {Artigas},\ and\ \citenamefont
  {Torner}}]{GomisBresco_Torner2017_BIC1D}%
  \BibitemOpen
  \bibfield  {author} {\bibinfo {author} {\bibfnamefont {J.}~\bibnamefont
  {Gomis-Bresco}}, \bibinfo {author} {\bibfnamefont {D.}~\bibnamefont
  {Artigas}},\ and\ \bibinfo {author} {\bibfnamefont {L.}~\bibnamefont
  {Torner}},\ }\bibfield  {title} {\bibinfo {title} {{Anisotropy-induced
  photonic bound states in the continuum}},\ }\href
  {https://doi.org/10.1038/nphoton.2017.31} {\bibfield  {journal} {\bibinfo
  {journal} {Nat. Photonics}\ }\textbf {\bibinfo {volume} {11}},\ \bibinfo
  {pages} {232} (\bibinfo {year} {2017})}\BibitemShut {NoStop}%
\bibitem [{\citenamefont {Mukherjee}\ \emph {et~al.}(2021)\citenamefont
  {Mukherjee}, \citenamefont {Gomis-Bresco}, \citenamefont {Artigas},\ and\
  \citenamefont {Torner}}]{Mukherjee2021}%
  \BibitemOpen
  \bibfield  {author} {\bibinfo {author} {\bibfnamefont {S.}~\bibnamefont
  {Mukherjee}}, \bibinfo {author} {\bibfnamefont {J.}~\bibnamefont
  {Gomis-Bresco}}, \bibinfo {author} {\bibfnamefont {D.}~\bibnamefont
  {Artigas}},\ and\ \bibinfo {author} {\bibfnamefont {L.}~\bibnamefont
  {Torner}},\ }\bibfield  {title} {\bibinfo {title} {{Unidirectional guided
  resonances in anisotropic waveguides}},\ }\href
  {https://doi.org/10.1364/OL.425393} {\bibfield  {journal} {\bibinfo
  {journal} {Optics Letters}\ }\textbf {\bibinfo {volume} {46}},\ \bibinfo
  {pages} {2545} (\bibinfo {year} {2021})}\BibitemShut {NoStop}%
\bibitem [{\citenamefont {Bulgakov}\ \emph {et~al.}(2022)\citenamefont
  {Bulgakov}, \citenamefont {Pilipchuk},\ and\ \citenamefont
  {Sadreev}}]{Bulgakov2022metalwaveguide}%
  \BibitemOpen
  \bibfield  {author} {\bibinfo {author} {\bibfnamefont {E.}~\bibnamefont
  {Bulgakov}}, \bibinfo {author} {\bibfnamefont {A.}~\bibnamefont
  {Pilipchuk}},\ and\ \bibinfo {author} {\bibfnamefont {A.}~\bibnamefont
  {Sadreev}},\ }\bibfield  {title} {\bibinfo {title} {Desktop laboratory of
  bound states in the continuum in metallic waveguide with dielectric
  cavities},\ }\href {https://doi.org/10.1103/PhysRevB.106.075304} {\bibfield
  {journal} {\bibinfo  {journal} {Phys. Rev. B}\ }\textbf {\bibinfo {volume}
  {106}},\ \bibinfo {pages} {075304} (\bibinfo {year} {2022})}\BibitemShut
  {NoStop}%
\bibitem [{\citenamefont {Rybin}\ \emph {et~al.}(2017)\citenamefont {Rybin},
  \citenamefont {Koshelev}, \citenamefont {Sadrieva}, \citenamefont {Samusev},
  \citenamefont {Bogdanov}, \citenamefont {Limonov},\ and\ \citenamefont
  {Kivshar}}]{Rybin2017}%
  \BibitemOpen
  \bibfield  {author} {\bibinfo {author} {\bibfnamefont {M.~V.}\ \bibnamefont
  {Rybin}}, \bibinfo {author} {\bibfnamefont {K.~L.}\ \bibnamefont {Koshelev}},
  \bibinfo {author} {\bibfnamefont {Z.~F.}\ \bibnamefont {Sadrieva}}, \bibinfo
  {author} {\bibfnamefont {K.~B.}\ \bibnamefont {Samusev}}, \bibinfo {author}
  {\bibfnamefont {A.~A.}\ \bibnamefont {Bogdanov}}, \bibinfo {author}
  {\bibfnamefont {M.~F.}\ \bibnamefont {Limonov}},\ and\ \bibinfo {author}
  {\bibfnamefont {Y.~S.}\ \bibnamefont {Kivshar}},\ }\bibfield  {title}
  {\bibinfo {title} {{High-Q Supercavity Modes in Subwavelength Dielectric
  Resonators}},\ }\href {https://doi.org/10.1103/PhysRevLett.119.243901}
  {\bibfield  {journal} {\bibinfo  {journal} {Phys. Rev. Lett.}\ }\textbf
  {\bibinfo {volume} {119}},\ \bibinfo {pages} {243901} (\bibinfo {year}
  {2017})}\BibitemShut {NoStop}%
\bibitem [{\citenamefont {Koshelev}\ \emph {et~al.}(2020)\citenamefont
  {Koshelev}, \citenamefont {Kruk}, \citenamefont {Melik-Gaykazyan},
  \citenamefont {Choi}, \citenamefont {Bogdanov}, \citenamefont {Park},\ and\
  \citenamefont {Kivshar}}]{Koshelev_nolinear2020}%
  \BibitemOpen
  \bibfield  {author} {\bibinfo {author} {\bibfnamefont {K.}~\bibnamefont
  {Koshelev}}, \bibinfo {author} {\bibfnamefont {S.}~\bibnamefont {Kruk}},
  \bibinfo {author} {\bibfnamefont {E.}~\bibnamefont {Melik-Gaykazyan}},
  \bibinfo {author} {\bibfnamefont {J.-H.}\ \bibnamefont {Choi}}, \bibinfo
  {author} {\bibfnamefont {A.}~\bibnamefont {Bogdanov}}, \bibinfo {author}
  {\bibfnamefont {H.-G.}\ \bibnamefont {Park}},\ and\ \bibinfo {author}
  {\bibfnamefont {Y.}~\bibnamefont {Kivshar}},\ }\bibfield  {title} {\bibinfo
  {title} {{Subwavelength dielectric resonators for nonlinear nanophotonics}},\
  }\href {https://doi.org/10.1126/science.aaz3985} {\bibfield  {journal}
  {\bibinfo  {journal} {Science}\ }\textbf {\bibinfo {volume} {367}},\ \bibinfo
  {pages} {288} (\bibinfo {year} {2020})}\BibitemShut {NoStop}%
\bibitem [{\citenamefont {Shi}\ \emph {et~al.}(2022)\citenamefont {Shi},
  \citenamefont {Deng}, \citenamefont {Geng}, \citenamefont {Zeng},
  \citenamefont {Zeng}, \citenamefont {Hu}, \citenamefont {Overvig},
  \citenamefont {Li}, \citenamefont {Qiu}, \citenamefont {Al{\`u}} \emph
  {et~al.}}]{shi2022chiral}%
  \BibitemOpen
  \bibfield  {author} {\bibinfo {author} {\bibfnamefont {T.}~\bibnamefont
  {Shi}}, \bibinfo {author} {\bibfnamefont {Z.-L.}\ \bibnamefont {Deng}},
  \bibinfo {author} {\bibfnamefont {G.}~\bibnamefont {Geng}}, \bibinfo {author}
  {\bibfnamefont {X.}~\bibnamefont {Zeng}}, \bibinfo {author} {\bibfnamefont
  {Y.}~\bibnamefont {Zeng}}, \bibinfo {author} {\bibfnamefont {G.}~\bibnamefont
  {Hu}}, \bibinfo {author} {\bibfnamefont {A.}~\bibnamefont {Overvig}},
  \bibinfo {author} {\bibfnamefont {J.}~\bibnamefont {Li}}, \bibinfo {author}
  {\bibfnamefont {C.-W.}\ \bibnamefont {Qiu}}, \bibinfo {author} {\bibfnamefont
  {A.}~\bibnamefont {Al{\`u}}}, \emph {et~al.},\ }\bibfield  {title} {\bibinfo
  {title} {Planar chiral metasurfaces with maximal and tunable chiroptical
  response driven by bound states in the continuum},\ }\href@noop {} {\bibfield
   {journal} {\bibinfo  {journal} {Nature Communications}\ }\textbf {\bibinfo
  {volume} {13}},\ \bibinfo {pages} {4111} (\bibinfo {year}
  {2022})}\BibitemShut {NoStop}%
\bibitem [{\citenamefont {Gorkunov}\ \emph {et~al.}(2020)\citenamefont
  {Gorkunov}, \citenamefont {Antonov},\ and\ \citenamefont
  {Kivshar}}]{gorkunov2020metasurfaces}%
  \BibitemOpen
  \bibfield  {author} {\bibinfo {author} {\bibfnamefont {M.~V.}\ \bibnamefont
  {Gorkunov}}, \bibinfo {author} {\bibfnamefont {A.~A.}\ \bibnamefont
  {Antonov}},\ and\ \bibinfo {author} {\bibfnamefont {Y.~S.}\ \bibnamefont
  {Kivshar}},\ }\bibfield  {title} {\bibinfo {title} {Metasurfaces with maximum
  chirality empowered by bound states in the continuum},\ }\href@noop {}
  {\bibfield  {journal} {\bibinfo  {journal} {Physical Review Letters}\
  }\textbf {\bibinfo {volume} {125}},\ \bibinfo {pages} {093903} (\bibinfo
  {year} {2020})}\BibitemShut {NoStop}%
\bibitem [{\citenamefont {Bulgakov}\ and\ \citenamefont
  {Sadreev}(2022)}]{Sadreev2022trapping}%
  \BibitemOpen
  \bibfield  {author} {\bibinfo {author} {\bibfnamefont {E.~N.}\ \bibnamefont
  {Bulgakov}}\ and\ \bibinfo {author} {\bibfnamefont {A.~F.}\ \bibnamefont
  {Sadreev}},\ }\bibfield  {title} {\bibinfo {title} {Self-trapping of
  nanoparticles by bound states in the continuum},\ }\href
  {https://doi.org/10.1103/PhysRevB.106.165430} {\bibfield  {journal} {\bibinfo
   {journal} {Phys. Rev. B}\ }\textbf {\bibinfo {volume} {106}},\ \bibinfo
  {pages} {165430} (\bibinfo {year} {2022})}\BibitemShut {NoStop}%
\bibitem [{\citenamefont {Wang}\ \emph {et~al.}(2023)\citenamefont {Wang},
  \citenamefont {Liang}, \citenamefont {Qu}, \citenamefont {Chen},
  \citenamefont {Cui}, \citenamefont {Cheng}, \citenamefont {Zhang},
  \citenamefont {Yao}, \citenamefont {Chen}, \citenamefont {Tsai} \emph
  {et~al.}}]{wang2023plasmonicBIC}%
  \BibitemOpen
  \bibfield  {author} {\bibinfo {author} {\bibfnamefont {Z.}~\bibnamefont
  {Wang}}, \bibinfo {author} {\bibfnamefont {Y.}~\bibnamefont {Liang}},
  \bibinfo {author} {\bibfnamefont {J.}~\bibnamefont {Qu}}, \bibinfo {author}
  {\bibfnamefont {M.~K.}\ \bibnamefont {Chen}}, \bibinfo {author}
  {\bibfnamefont {M.}~\bibnamefont {Cui}}, \bibinfo {author} {\bibfnamefont
  {Z.}~\bibnamefont {Cheng}}, \bibinfo {author} {\bibfnamefont
  {J.}~\bibnamefont {Zhang}}, \bibinfo {author} {\bibfnamefont
  {J.}~\bibnamefont {Yao}}, \bibinfo {author} {\bibfnamefont {S.}~\bibnamefont
  {Chen}}, \bibinfo {author} {\bibfnamefont {D.~P.}\ \bibnamefont {Tsai}},
  \emph {et~al.},\ }\bibfield  {title} {\bibinfo {title} {Plasmonic bound
  states in the continuum for unpolarized weak spatially coherent light},\
  }\href@noop {} {\bibfield  {journal} {\bibinfo  {journal} {Photonics
  Research}\ }\textbf {\bibinfo {volume} {11}},\ \bibinfo {pages} {260}
  (\bibinfo {year} {2023})}\BibitemShut {NoStop}%
\bibitem [{\citenamefont {Bikbaev}\ \emph {et~al.}(2023)\citenamefont
  {Bikbaev}, \citenamefont {Maksimov}, \citenamefont {Pankin}, \citenamefont
  {Ye}, \citenamefont {Chen},\ and\ \citenamefont
  {Timofeev}}]{bikbaev2023absBIC}%
  \BibitemOpen
  \bibfield  {author} {\bibinfo {author} {\bibfnamefont {R.~G.}\ \bibnamefont
  {Bikbaev}}, \bibinfo {author} {\bibfnamefont {D.~N.}\ \bibnamefont
  {Maksimov}}, \bibinfo {author} {\bibfnamefont {P.~S.}\ \bibnamefont
  {Pankin}}, \bibinfo {author} {\bibfnamefont {M.-J.}\ \bibnamefont {Ye}},
  \bibinfo {author} {\bibfnamefont {K.-P.}\ \bibnamefont {Chen}},\ and\
  \bibinfo {author} {\bibfnamefont {I.~V.}\ \bibnamefont {Timofeev}},\
  }\bibfield  {title} {\bibinfo {title} {Enhanced light absorption in tamm
  metasurface with a bound state in the continuum},\ }\href@noop {} {\bibfield
  {journal} {\bibinfo  {journal} {Photonics and Nanostructures-Fundamentals and
  Applications}\ ,\ \bibinfo {pages} {101148}} (\bibinfo {year}
  {2023})}\BibitemShut {NoStop}%
\bibitem [{\citenamefont {Romano}\ \emph {et~al.}(2020)\citenamefont {Romano},
  \citenamefont {Mangini}, \citenamefont {Penzo}, \citenamefont {Cabrini},
  \citenamefont {De~Luca}, \citenamefont {Rendina}, \citenamefont {Mocella},\
  and\ \citenamefont {Zito}}]{romano2020ultrasensitive}%
  \BibitemOpen
  \bibfield  {author} {\bibinfo {author} {\bibfnamefont {S.}~\bibnamefont
  {Romano}}, \bibinfo {author} {\bibfnamefont {M.}~\bibnamefont {Mangini}},
  \bibinfo {author} {\bibfnamefont {E.}~\bibnamefont {Penzo}}, \bibinfo
  {author} {\bibfnamefont {S.}~\bibnamefont {Cabrini}}, \bibinfo {author}
  {\bibfnamefont {A.~C.}\ \bibnamefont {De~Luca}}, \bibinfo {author}
  {\bibfnamefont {I.}~\bibnamefont {Rendina}}, \bibinfo {author} {\bibfnamefont
  {V.}~\bibnamefont {Mocella}},\ and\ \bibinfo {author} {\bibfnamefont
  {G.}~\bibnamefont {Zito}},\ }\bibfield  {title} {\bibinfo {title}
  {Ultrasensitive surface refractive index imaging based on quasi-bound states
  in the continuum},\ }\href@noop {} {\bibfield  {journal} {\bibinfo  {journal}
  {ACS nano}\ }\textbf {\bibinfo {volume} {14}},\ \bibinfo {pages} {15417}
  (\bibinfo {year} {2020})}\BibitemShut {NoStop}%
\bibitem [{\citenamefont {Maksimov}\ \emph {et~al.}(2020)\citenamefont
  {Maksimov}, \citenamefont {Gerasimov}, \citenamefont {Romano},\ and\
  \citenamefont {Polyutov}}]{Maksimov2020sensor}%
  \BibitemOpen
  \bibfield  {author} {\bibinfo {author} {\bibfnamefont {D.~N.}\ \bibnamefont
  {Maksimov}}, \bibinfo {author} {\bibfnamefont {V.~S.}\ \bibnamefont
  {Gerasimov}}, \bibinfo {author} {\bibfnamefont {S.}~\bibnamefont {Romano}},\
  and\ \bibinfo {author} {\bibfnamefont {S.~P.}\ \bibnamefont {Polyutov}},\
  }\bibfield  {title} {\bibinfo {title} {Refractive index sensing with optical
  bound states in the continuum},\ }\href {https://doi.org/10.1364/OE.411749}
  {\bibfield  {journal} {\bibinfo  {journal} {Opt. Express}\ }\textbf {\bibinfo
  {volume} {28}},\ \bibinfo {pages} {38907} (\bibinfo {year}
  {2020})}\BibitemShut {NoStop}%
\bibitem [{\citenamefont {Yusupov}\ \emph {et~al.}(2021)\citenamefont
  {Yusupov}, \citenamefont {Filonov}, \citenamefont {Bogdanov}, \citenamefont
  {Ginzburg}, \citenamefont {Rybin},\ and\ \citenamefont
  {Slobozhanyuk}}]{Yusupov2021}%
  \BibitemOpen
  \bibfield  {author} {\bibinfo {author} {\bibfnamefont {I.}~\bibnamefont
  {Yusupov}}, \bibinfo {author} {\bibfnamefont {D.}~\bibnamefont {Filonov}},
  \bibinfo {author} {\bibfnamefont {A.}~\bibnamefont {Bogdanov}}, \bibinfo
  {author} {\bibfnamefont {P.}~\bibnamefont {Ginzburg}}, \bibinfo {author}
  {\bibfnamefont {M.~V.}\ \bibnamefont {Rybin}},\ and\ \bibinfo {author}
  {\bibfnamefont {A.}~\bibnamefont {Slobozhanyuk}},\ }\bibfield  {title}
  {\bibinfo {title} {{Chipless wireless temperature sensor based on quasi-BIC
  resonance}},\ }\href {https://doi.org/10.1063/5.0064480} {\bibfield
  {journal} {\bibinfo  {journal} {Applied Physics Letters}\ }\textbf {\bibinfo
  {volume} {119}},\ \bibinfo {pages} {193504} (\bibinfo {year}
  {2021})}\BibitemShut {NoStop}%
\bibitem [{\citenamefont {Chernyak}\ \emph {et~al.}(2020)\citenamefont
  {Chernyak}, \citenamefont {Barsukova}, \citenamefont {Shorokhov},
  \citenamefont {Musorin},\ and\ \citenamefont
  {Fedyanin}}]{Fedyanin2022sensor}%
  \BibitemOpen
  \bibfield  {author} {\bibinfo {author} {\bibfnamefont {A.~M.}\ \bibnamefont
  {Chernyak}}, \bibinfo {author} {\bibfnamefont {M.~G.}\ \bibnamefont
  {Barsukova}}, \bibinfo {author} {\bibfnamefont {A.~S.}\ \bibnamefont
  {Shorokhov}}, \bibinfo {author} {\bibfnamefont {A.~I.}\ \bibnamefont
  {Musorin}},\ and\ \bibinfo {author} {\bibfnamefont {A.~A.}\ \bibnamefont
  {Fedyanin}},\ }\bibfield  {title} {\bibinfo {title} {{Bound States in the
  Continuum in Magnetophotonic Metasurfaces}},\ }\href
  {https://doi.org/10.1134/S0021364020010105} {\bibfield  {journal} {\bibinfo
  {journal} {JETP Letters}\ }\textbf {\bibinfo {volume} {111}},\ \bibinfo
  {pages} {46} (\bibinfo {year} {2020})}\BibitemShut {NoStop}%
\bibitem [{\citenamefont {Schiattarella}\ \emph {et~al.}(2022)\citenamefont
  {Schiattarella}, \citenamefont {Sanit{\`{a}}}, \citenamefont {Alulema},
  \citenamefont {Lanzio}, \citenamefont {Cabrini}, \citenamefont {Lamberti},
  \citenamefont {Rendina}, \citenamefont {Mocella}, \citenamefont {Zito},\ and\
  \citenamefont {Romano}}]{Romano2022_BICSensor}%
  \BibitemOpen
  \bibfield  {author} {\bibinfo {author} {\bibfnamefont {C.}~\bibnamefont
  {Schiattarella}}, \bibinfo {author} {\bibfnamefont {G.}~\bibnamefont
  {Sanit{\`{a}}}}, \bibinfo {author} {\bibfnamefont {B.~G.}\ \bibnamefont
  {Alulema}}, \bibinfo {author} {\bibfnamefont {V.}~\bibnamefont {Lanzio}},
  \bibinfo {author} {\bibfnamefont {S.}~\bibnamefont {Cabrini}}, \bibinfo
  {author} {\bibfnamefont {A.}~\bibnamefont {Lamberti}}, \bibinfo {author}
  {\bibfnamefont {I.}~\bibnamefont {Rendina}}, \bibinfo {author} {\bibfnamefont
  {V.}~\bibnamefont {Mocella}}, \bibinfo {author} {\bibfnamefont
  {G.}~\bibnamefont {Zito}},\ and\ \bibinfo {author} {\bibfnamefont
  {S.}~\bibnamefont {Romano}},\ }\bibfield  {title} {\bibinfo {title} {{High-Q
  photonic aptasensor based on avoided crossing bound states in the continuum
  and trace detection of ochratoxin A}},\ }\href
  {https://doi.org/10.1016/j.biosx.2022.100262} {\bibfield  {journal} {\bibinfo
   {journal} {Biosensors and Bioelectronics: X}\ ,\ \bibinfo {pages} {100262}}
  (\bibinfo {year} {2022})}\BibitemShut {NoStop}%
\bibitem [{\citenamefont {Sadreev}\ \emph {et~al.}(2022)\citenamefont
  {Sadreev}, \citenamefont {Bulgakov}, \citenamefont {Pilipchuk}, \citenamefont
  {Miroshnichenko},\ and\ \citenamefont {Huang}}]{Sadreev2022PRB}%
  \BibitemOpen
  \bibfield  {author} {\bibinfo {author} {\bibfnamefont {A.}~\bibnamefont
  {Sadreev}}, \bibinfo {author} {\bibfnamefont {E.}~\bibnamefont {Bulgakov}},
  \bibinfo {author} {\bibfnamefont {A.}~\bibnamefont {Pilipchuk}}, \bibinfo
  {author} {\bibfnamefont {A.}~\bibnamefont {Miroshnichenko}},\ and\ \bibinfo
  {author} {\bibfnamefont {L.}~\bibnamefont {Huang}},\ }\bibfield  {title}
  {\bibinfo {title} {{Degenerate bound states in the continuum in square and
  triangular open acoustic resonators}},\ }\href
  {https://doi.org/10.1103/PhysRevB.106.085404} {\bibfield  {journal} {\bibinfo
   {journal} {Physical Review B}\ }\textbf {\bibinfo {volume} {106}},\ \bibinfo
  {pages} {085404} (\bibinfo {year} {2022})}\BibitemShut {NoStop}%
\bibitem [{\citenamefont {Maksimov}\ \emph {et~al.}(2022)\citenamefont
  {Maksimov}, \citenamefont {Kostyukov}, \citenamefont {Ershov}, \citenamefont
  {Molokeev}, \citenamefont {Bulgakov},\ and\ \citenamefont
  {Gerasimov}}]{Maksimov2022}%
  \BibitemOpen
  \bibfield  {author} {\bibinfo {author} {\bibfnamefont {D.~N.}\ \bibnamefont
  {Maksimov}}, \bibinfo {author} {\bibfnamefont {A.~S.}\ \bibnamefont
  {Kostyukov}}, \bibinfo {author} {\bibfnamefont {A.~E.}\ \bibnamefont
  {Ershov}}, \bibinfo {author} {\bibfnamefont {M.~S.}\ \bibnamefont
  {Molokeev}}, \bibinfo {author} {\bibfnamefont {E.~N.}\ \bibnamefont
  {Bulgakov}},\ and\ \bibinfo {author} {\bibfnamefont {V.~S.}\ \bibnamefont
  {Gerasimov}},\ }\bibfield  {title} {\bibinfo {title} {{Thermo-optic
  hysteresis with bound states in the continuum}},\ }\href
  {http://arxiv.org/abs/2210.02364} {\bibfield  {journal} {\bibinfo  {journal}
  {arXiv Prepr.}\ } (\bibinfo {year} {2022})},\ \Eprint
  {https://arxiv.org/abs/2210.02364} {arXiv:2210.02364} \BibitemShut {NoStop}%
\bibitem [{\citenamefont {Wu}\ \emph {et~al.}(2019)\citenamefont {Wu},
  \citenamefont {Wu}, \citenamefont {Guo}, \citenamefont {Jiang}, \citenamefont
  {Sun}, \citenamefont {Li}, \citenamefont {Ren},\ and\ \citenamefont
  {Chen}}]{wu2019giantGH_BIC}%
  \BibitemOpen
  \bibfield  {author} {\bibinfo {author} {\bibfnamefont {F.}~\bibnamefont
  {Wu}}, \bibinfo {author} {\bibfnamefont {J.}~\bibnamefont {Wu}}, \bibinfo
  {author} {\bibfnamefont {Z.}~\bibnamefont {Guo}}, \bibinfo {author}
  {\bibfnamefont {H.}~\bibnamefont {Jiang}}, \bibinfo {author} {\bibfnamefont
  {Y.}~\bibnamefont {Sun}}, \bibinfo {author} {\bibfnamefont {Y.}~\bibnamefont
  {Li}}, \bibinfo {author} {\bibfnamefont {J.}~\bibnamefont {Ren}},\ and\
  \bibinfo {author} {\bibfnamefont {H.}~\bibnamefont {Chen}},\ }\bibfield
  {title} {\bibinfo {title} {Giant enhancement of the goos-hanchen shift
  assisted by quasibound states in the continuum},\ }\href@noop {} {\bibfield
  {journal} {\bibinfo  {journal} {Physical Review Applied}\ }\textbf {\bibinfo
  {volume} {12}},\ \bibinfo {pages} {014028} (\bibinfo {year}
  {2019})}\BibitemShut {NoStop}%
\bibitem [{\citenamefont {Yusupov}\ \emph {et~al.}(2022)\citenamefont
  {Yusupov}, \citenamefont {Filonov}, \citenamefont {Bogdanov}, \citenamefont
  {Ginzburg}, \citenamefont {Rybin},\ and\ \citenamefont
  {Slobozhanyuk}}]{Yusupov2020tempBICsensor}%
  \BibitemOpen
  \bibfield  {author} {\bibinfo {author} {\bibfnamefont {I.}~\bibnamefont
  {Yusupov}}, \bibinfo {author} {\bibfnamefont {D.}~\bibnamefont {Filonov}},
  \bibinfo {author} {\bibfnamefont {A.}~\bibnamefont {Bogdanov}}, \bibinfo
  {author} {\bibfnamefont {P.}~\bibnamefont {Ginzburg}}, \bibinfo {author}
  {\bibfnamefont {M.~V.}\ \bibnamefont {Rybin}},\ and\ \bibinfo {author}
  {\bibfnamefont {A.}~\bibnamefont {Slobozhanyuk}},\ }\bibfield  {title}
  {\bibinfo {title} {Passive temperature sensor tag based on quasi-bic},\ }in\
  \href {https://doi.org/10.23919/SpliTech55088.2022.9854305} {\emph {\bibinfo
  {booktitle} {2022 7th International Conference on Smart and Sustainable
  Technologies (SpliTech)}}}\ (\bibinfo {year} {2022})\ pp.\ \bibinfo {pages}
  {1--4}\BibitemShut {NoStop}%
\bibitem [{\citenamefont {Arkhipkin}\ \emph {et~al.}(2008)\citenamefont
  {Arkhipkin}, \citenamefont {Gunyakov}, \citenamefont {Myslivets},
  \citenamefont {Gerasimov}, \citenamefont {Zyryanov}, \citenamefont {Vetrov},\
  and\ \citenamefont {Shabanov}}]{Arkhipkin2008}%
  \BibitemOpen
  \bibfield  {author} {\bibinfo {author} {\bibfnamefont {V.~G.}\ \bibnamefont
  {Arkhipkin}}, \bibinfo {author} {\bibfnamefont {V.~A.}\ \bibnamefont
  {Gunyakov}}, \bibinfo {author} {\bibfnamefont {S.~A.}\ \bibnamefont
  {Myslivets}}, \bibinfo {author} {\bibfnamefont {V.~P.}\ \bibnamefont
  {Gerasimov}}, \bibinfo {author} {\bibfnamefont {V.~Y.}\ \bibnamefont
  {Zyryanov}}, \bibinfo {author} {\bibfnamefont {S.~Y.}\ \bibnamefont
  {Vetrov}},\ and\ \bibinfo {author} {\bibfnamefont {V.~F.}\ \bibnamefont
  {Shabanov}},\ }\bibfield  {title} {\bibinfo {title} {{One-dimensional
  photonic crystals with a planar oriented nematic layer: Temperature and
  angular dependence of the spectra of defect modes}},\ }\href
  {https://doi.org/10.1007/s11447-008-2017-9} {\bibfield  {journal} {\bibinfo
  {journal} {J. Exp. Theor. Phys.}\ }\textbf {\bibinfo {volume} {106}},\
  \bibinfo {pages} {388} (\bibinfo {year} {2008})}\BibitemShut {NoStop}%
\bibitem [{\citenamefont {Pankin}\ \emph {et~al.}(2021)\citenamefont {Pankin},
  \citenamefont {Sutormin}, \citenamefont {Gunyakov}, \citenamefont {Zelenov},
  \citenamefont {Tambasov}, \citenamefont {Masyugin}, \citenamefont
  {Volochaev}, \citenamefont {Baron}, \citenamefont {Chen}, \citenamefont
  {Zyryanov}, \citenamefont {Vetrov},\ and\ \citenamefont
  {Timofeev}}]{Pankin2021APL}%
  \BibitemOpen
  \bibfield  {author} {\bibinfo {author} {\bibfnamefont {P.~S.}\ \bibnamefont
  {Pankin}}, \bibinfo {author} {\bibfnamefont {V.~S.}\ \bibnamefont
  {Sutormin}}, \bibinfo {author} {\bibfnamefont {V.~A.}\ \bibnamefont
  {Gunyakov}}, \bibinfo {author} {\bibfnamefont {F.~V.}\ \bibnamefont
  {Zelenov}}, \bibinfo {author} {\bibfnamefont {I.~A.}\ \bibnamefont
  {Tambasov}}, \bibinfo {author} {\bibfnamefont {A.~N.}\ \bibnamefont
  {Masyugin}}, \bibinfo {author} {\bibfnamefont {M.~N.}\ \bibnamefont
  {Volochaev}}, \bibinfo {author} {\bibfnamefont {F.~A.}\ \bibnamefont
  {Baron}}, \bibinfo {author} {\bibfnamefont {K.~P.}\ \bibnamefont {Chen}},
  \bibinfo {author} {\bibfnamefont {V.~Y.}\ \bibnamefont {Zyryanov}}, \bibinfo
  {author} {\bibfnamefont {S.~Y.}\ \bibnamefont {Vetrov}},\ and\ \bibinfo
  {author} {\bibfnamefont {I.~V.}\ \bibnamefont {Timofeev}},\ }\bibfield
  {title} {\bibinfo {title} {{Experimental implementation of tunable hybrid
  Tamm-microcavity modes}},\ }\href {https://doi.org/10.1063/5.0067179}
  {\bibfield  {journal} {\bibinfo  {journal} {Applied Physics Letters}\
  }\textbf {\bibinfo {volume} {119}},\ \bibinfo {pages} {161107} (\bibinfo
  {year} {2021})}\BibitemShut {NoStop}%
\bibitem [{\citenamefont {Buzin}\ \emph {et~al.}(2023)\citenamefont {Buzin},
  \citenamefont {Pankin}, \citenamefont {Maksimov}, \citenamefont {Romanenko},
  \citenamefont {Sutormin}, \citenamefont {Nabol}, \citenamefont {Zelenov},
  \citenamefont {Masyugin}, \citenamefont {Volochaev}, \citenamefont {Vetrov}
  \emph {et~al.}}]{buzin2023hybrid}%
  \BibitemOpen
  \bibfield  {author} {\bibinfo {author} {\bibfnamefont {D.}~\bibnamefont
  {Buzin}}, \bibinfo {author} {\bibfnamefont {P.}~\bibnamefont {Pankin}},
  \bibinfo {author} {\bibfnamefont {D.}~\bibnamefont {Maksimov}}, \bibinfo
  {author} {\bibfnamefont {G.}~\bibnamefont {Romanenko}}, \bibinfo {author}
  {\bibfnamefont {V.}~\bibnamefont {Sutormin}}, \bibinfo {author}
  {\bibfnamefont {S.}~\bibnamefont {Nabol}}, \bibinfo {author} {\bibfnamefont
  {F.}~\bibnamefont {Zelenov}}, \bibinfo {author} {\bibfnamefont
  {A.}~\bibnamefont {Masyugin}}, \bibinfo {author} {\bibfnamefont
  {M.}~\bibnamefont {Volochaev}}, \bibinfo {author} {\bibfnamefont {S.~Y.}\
  \bibnamefont {Vetrov}}, \emph {et~al.},\ }\bibfield  {title} {\bibinfo
  {title} {Hybrid tamm and quasi-bic microcavity modes},\ }\href@noop {}
  {\bibfield  {journal} {\bibinfo  {journal} {arXiv preprint arXiv:2306.08455}\
  } (\bibinfo {year} {2023})}\BibitemShut {NoStop}%
\bibitem [{\citenamefont {Blinov}(2010)}]{Blinov2010bk}%
  \BibitemOpen
  \bibfield  {author} {\bibinfo {author} {\bibfnamefont {L.~M.}\ \bibnamefont
  {Blinov}},\ }\href@noop {} {\emph {\bibinfo {title} {{Structure and
  Properties of Liquid Crystals}}}},\ Topics in applied physics\ (\bibinfo
  {publisher} {Springer},\ \bibinfo {year} {2010})\ p.\ \bibinfo {pages}
  {458}\BibitemShut {NoStop}%
\bibitem [{\citenamefont {Akhmanov}\ and\ \citenamefont
  {Nikitin}(1997)}]{Akhmanov1997bk}%
  \BibitemOpen
  \bibfield  {author} {\bibinfo {author} {\bibfnamefont {S.~A.}\ \bibnamefont
  {Akhmanov}}\ and\ \bibinfo {author} {\bibfnamefont {S.~Y.}\ \bibnamefont
  {Nikitin}},\ }\href@noop {} {\emph {\bibinfo {title} {{Physical Optics}}}}\
  (\bibinfo  {publisher} {Clarendon Press},\ \bibinfo {year}
  {1997})\BibitemShut {NoStop}%
\bibitem [{\citenamefont {Ignatovich}\ and\ \citenamefont
  {Ignatovich}(2012)}]{Ignatovich2012}%
  \BibitemOpen
  \bibfield  {author} {\bibinfo {author} {\bibfnamefont {F.~V.}\ \bibnamefont
  {Ignatovich}}\ and\ \bibinfo {author} {\bibfnamefont {V.~K.}\ \bibnamefont
  {Ignatovich}},\ }\bibfield  {title} {\bibinfo {title} {{Optics of anisotropic
  media}},\ }\href {https://doi.org/10.3367/UFNr.0182.201207f.0759} {\bibfield
  {journal} {\bibinfo  {journal} {Uspekhi Fiz. Nauk}\ }\textbf {\bibinfo
  {volume} {182}},\ \bibinfo {pages} {759} (\bibinfo {year}
  {2012})}\BibitemShut {NoStop}%
\bibitem [{\citenamefont {Friedrich}\ and\ \citenamefont
  {Wintgen}(1985)}]{Friedrich1985}%
  \BibitemOpen
  \bibfield  {author} {\bibinfo {author} {\bibfnamefont {H.}~\bibnamefont
  {Friedrich}}\ and\ \bibinfo {author} {\bibfnamefont {D.}~\bibnamefont
  {Wintgen}},\ }\bibfield  {title} {\bibinfo {title} {{Interfering resonances
  and bound states in the continuum}},\ }\href
  {https://doi.org/10.1103/PhysRevA.32.3231} {\bibfield  {journal} {\bibinfo
  {journal} {Phys. Rev. A}\ }\textbf {\bibinfo {volume} {32}},\ \bibinfo
  {pages} {3231} (\bibinfo {year} {1985})}\BibitemShut {NoStop}%
\bibitem [{\citenamefont {Berreman}(1972)}]{Berreman1972}%
  \BibitemOpen
  \bibfield  {author} {\bibinfo {author} {\bibfnamefont {D.~W.}\ \bibnamefont
  {Berreman}},\ }\bibfield  {title} {\bibinfo {title} {{Optics in stratified
  and anisotropic media: 4$\times$ 4-matrix formulation}},\ }\href@noop {}
  {\bibfield  {journal} {\bibinfo  {journal} {Journal of Optical Society of
  America}\ }\textbf {\bibinfo {volume} {62}},\ \bibinfo {pages} {502}
  (\bibinfo {year} {1972})}\BibitemShut {NoStop}%
\bibitem [{\citenamefont {Luke}\ \emph {et~al.}(2015)\citenamefont {Luke},
  \citenamefont {Okawachi}, \citenamefont {Lamont}, \citenamefont {Gaeta},\
  and\ \citenamefont {Lipson}}]{luke2015broadband}%
  \BibitemOpen
  \bibfield  {author} {\bibinfo {author} {\bibfnamefont {K.}~\bibnamefont
  {Luke}}, \bibinfo {author} {\bibfnamefont {Y.}~\bibnamefont {Okawachi}},
  \bibinfo {author} {\bibfnamefont {M.~R.~E.}\ \bibnamefont {Lamont}}, \bibinfo
  {author} {\bibfnamefont {A.~L.}\ \bibnamefont {Gaeta}},\ and\ \bibinfo
  {author} {\bibfnamefont {M.}~\bibnamefont {Lipson}},\ }\bibfield  {title}
  {\bibinfo {title} {{Broadband mid-infrared frequency comb generation in a Si
  3 N 4 microresonator}},\ }\href@noop {} {\bibfield  {journal} {\bibinfo
  {journal} {Optics letters}\ }\textbf {\bibinfo {volume} {40}},\ \bibinfo
  {pages} {4823} (\bibinfo {year} {2015})}\BibitemShut {NoStop}%
\bibitem [{\citenamefont {Gao}\ \emph {et~al.}(2013)\citenamefont {Gao},
  \citenamefont {Lemarchand},\ and\ \citenamefont {Lequime}}]{Gao2013_RI_SIO2}%
  \BibitemOpen
  \bibfield  {author} {\bibinfo {author} {\bibfnamefont {L.}~\bibnamefont
  {Gao}}, \bibinfo {author} {\bibfnamefont {F.}~\bibnamefont {Lemarchand}},\
  and\ \bibinfo {author} {\bibfnamefont {M.}~\bibnamefont {Lequime}},\
  }\bibfield  {title} {\bibinfo {title} {{Refractive index determination of
  SiO2 layer in the UV/Vis/NIR range: spectrophotometric reverse engineering on
  single and bi-layer designs}},\ }\href@noop {} {\bibfield  {journal}
  {\bibinfo  {journal} {Journal of the European Optical Society-Rapid
  publications}\ }\textbf {\bibinfo {volume} {8}} (\bibinfo {year}
  {2013})}\BibitemShut {NoStop}%
\bibitem [{\citenamefont {Schnepf}\ \emph {et~al.}(2017)\citenamefont
  {Schnepf}, \citenamefont {Mayer}, \citenamefont {Kuttner}, \citenamefont
  {Tebbe}, \citenamefont {Wolf}, \citenamefont {Dulle}, \citenamefont
  {Altantzis}, \citenamefont {Formanek}, \citenamefont {F{\"{o}}rster},
  \citenamefont {Bals}, \citenamefont {K{\"{o}}nig},\ and\ \citenamefont
  {Fery}}]{Schnepf2017PVA_RI}%
  \BibitemOpen
  \bibfield  {author} {\bibinfo {author} {\bibfnamefont {M.~J.}\ \bibnamefont
  {Schnepf}}, \bibinfo {author} {\bibfnamefont {M.}~\bibnamefont {Mayer}},
  \bibinfo {author} {\bibfnamefont {C.}~\bibnamefont {Kuttner}}, \bibinfo
  {author} {\bibfnamefont {M.}~\bibnamefont {Tebbe}}, \bibinfo {author}
  {\bibfnamefont {D.}~\bibnamefont {Wolf}}, \bibinfo {author} {\bibfnamefont
  {M.}~\bibnamefont {Dulle}}, \bibinfo {author} {\bibfnamefont
  {T.}~\bibnamefont {Altantzis}}, \bibinfo {author} {\bibfnamefont
  {P.}~\bibnamefont {Formanek}}, \bibinfo {author} {\bibfnamefont
  {S.}~\bibnamefont {F{\"{o}}rster}}, \bibinfo {author} {\bibfnamefont
  {S.}~\bibnamefont {Bals}}, \bibinfo {author} {\bibfnamefont {T.~A.~F.}\
  \bibnamefont {K{\"{o}}nig}},\ and\ \bibinfo {author} {\bibfnamefont
  {A.}~\bibnamefont {Fery}},\ }\bibfield  {title} {\bibinfo {title}
  {{Nanorattles with tailored electric field enhancement}},\ }\href
  {https://doi.org/10.1039/C7NR02952G} {\bibfield  {journal} {\bibinfo
  {journal} {Nanoscale}\ }\textbf {\bibinfo {volume} {9}},\ \bibinfo {pages}
  {9376} (\bibinfo {year} {2017})}\BibitemShut {NoStop}%
\bibitem [{\citenamefont {Tambasov}\ \emph {et~al.}(2019)\citenamefont
  {Tambasov}, \citenamefont {Volochaev}, \citenamefont {Voronin}, \citenamefont
  {Evsevskaya}, \citenamefont {Masyugin}, \citenamefont {Aleksandrovskii},
  \citenamefont {Smolyarova}, \citenamefont {Nemtsev}, \citenamefont
  {Lyashchenko}, \citenamefont {Bondarenko},\ and\ \citenamefont
  {Others}}]{tambasov2019structural}%
  \BibitemOpen
  \bibfield  {author} {\bibinfo {author} {\bibfnamefont {I.~A.}\ \bibnamefont
  {Tambasov}}, \bibinfo {author} {\bibfnamefont {M.~N.}\ \bibnamefont
  {Volochaev}}, \bibinfo {author} {\bibfnamefont {A.~S.}\ \bibnamefont
  {Voronin}}, \bibinfo {author} {\bibfnamefont {N.~P.}\ \bibnamefont
  {Evsevskaya}}, \bibinfo {author} {\bibfnamefont {A.~N.}\ \bibnamefont
  {Masyugin}}, \bibinfo {author} {\bibfnamefont {A.~S.}\ \bibnamefont
  {Aleksandrovskii}}, \bibinfo {author} {\bibfnamefont {T.~E.}\ \bibnamefont
  {Smolyarova}}, \bibinfo {author} {\bibfnamefont {I.~V.}\ \bibnamefont
  {Nemtsev}}, \bibinfo {author} {\bibfnamefont {S.~A.}\ \bibnamefont
  {Lyashchenko}}, \bibinfo {author} {\bibfnamefont {G.~N.}\ \bibnamefont
  {Bondarenko}},\ and\ \bibinfo {author} {\bibnamefont {Others}},\ }\bibfield
  {title} {\bibinfo {title} {{Structural, Optical, and Thermoelectric
  Properties of the ZnO: Al Films Synthesized by Atomic Layer Deposition}},\
  }\href@noop {} {\bibfield  {journal} {\bibinfo  {journal} {Physics of the
  Solid State}\ }\textbf {\bibinfo {volume} {61}},\ \bibinfo {pages} {1904}
  (\bibinfo {year} {2019})}\BibitemShut {NoStop}%
\bibitem [{\citenamefont {Wu}\ \emph {et~al.}(1993)\citenamefont {Wu},
  \citenamefont {Wu}, \citenamefont {Warenghem},\ and\ \citenamefont
  {Ismaili}}]{wu1993refractive}%
  \BibitemOpen
  \bibfield  {author} {\bibinfo {author} {\bibfnamefont {S.-T.}\ \bibnamefont
  {Wu}}, \bibinfo {author} {\bibfnamefont {C.}~\bibnamefont {Wu}}, \bibinfo
  {author} {\bibfnamefont {M.}~\bibnamefont {Warenghem}},\ and\ \bibinfo
  {author} {\bibfnamefont {M.}~\bibnamefont {Ismaili}},\ }\bibfield  {title}
  {\bibinfo {title} {Refractive index dispersions of liquid crystals},\
  }\href@noop {} {\bibfield  {journal} {\bibinfo  {journal} {Optical
  Engineering}\ }\textbf {\bibinfo {volume} {32}},\ \bibinfo {pages} {1775}
  (\bibinfo {year} {1993})}\BibitemShut {NoStop}%
\bibitem [{\citenamefont {Fan}\ \emph {et~al.}(2003)\citenamefont {Fan},
  \citenamefont {Suh},\ and\ \citenamefont {Joannopoulos}}]{FanShanhui2003}%
  \BibitemOpen
  \bibfield  {author} {\bibinfo {author} {\bibfnamefont {S.}~\bibnamefont
  {Fan}}, \bibinfo {author} {\bibfnamefont {W.}~\bibnamefont {Suh}},\ and\
  \bibinfo {author} {\bibfnamefont {J.~D.}\ \bibnamefont {Joannopoulos}},\
  }\bibfield  {title} {\bibinfo {title} {{Temporal coupled-mode theory for the
  Fano resonance in optical resonators}},\ }\href
  {https://doi.org/10.1364/JOSAA.20.000569} {\bibfield  {journal} {\bibinfo
  {journal} {J. Opt. Soc. Am. A}\ }\textbf {\bibinfo {volume} {20}},\ \bibinfo
  {pages} {569} (\bibinfo {year} {2003})}\BibitemShut {NoStop}%
\end{thebibliography}%

\end{document}